\pdfoutput=1
\documentclass[aps,pre,reprint,groupedaddress,nofootinbib,twocolumn]{revtex4-2}
\usepackage{acronym}
\usepackage[dvipsnames]{xcolor}
\usepackage{amsmath,amssymb}
\usepackage{graphicx}
\usepackage{grffile} 
\usepackage{footmisc}
\usepackage{bbold}
\usepackage{hyperref}

\newcommand{\Er}{E_\mathrm{r}}
\newcommand{\Eth}{E_\mathrm{th}}
\newcommand{\Eis}{E_\mathrm{IS}}
\newcommand{\Emb}{E_\mathrm{MB}}
\newcommand{\tmb}{\tau_\mathrm{MB}}
\newcommand{\tis}{\tau_\mathrm{IS}}
\newcommand{\ta}{\tau_\alpha}
\newcommand{\qmb}{q_\mathrm{MB}}
\newcommand{\qis}{q_\mathrm{IS}}
\newcommand{\np}{n_\mathrm{pivots}}
\newcommand{\Td}{T_\mathrm{d}}
\newcommand{\To}{T_\mathrm{o}}

\graphicspath{{./figures/}{./}}

\begin{document}
\title{Revisiting the Concept of Activation in Supercooled Liquids}
\author{Marco Baity-Jesi$^{1}$, Giulio Biroli$^2$, David R. Reichman$^3$}
\affiliation{$^1$ Eawag, \"Uberlandstrasse 133, 8600 D\"ubendorf, Switzerland\,}
\email[]{marco.baityjesi@eawag.ch}
\affiliation{$^2$ Departement de Physique Statistique, \'Ecole Normale Sup\'erieure, 75005 Paris, France}
\affiliation{$^3$ Department of Chemistry, Columbia University, New York, NY 10027, USA}

\date{\today}

\begin{abstract}
In this work we revisit the description of dynamics based on the concepts of metabasins and activation
in mildly supercooled liquids via the analysis of the dynamics of a paradigmatic glass former 
between its onset temperature $T_{o}$ and mode-coupling temperature $T_{c}$.
First, we provide measures that demonstrate that the onset of glassiness is indeed connected to the landscape, 
and that metabasin waiting time distributions are so broad that the system can 
remain stuck in a metabasin for times that exceed $\ta$ by orders of magnitude.
We then reanalyze the transitions between metabasins, providing several indications that the standard picture of activated dynamics in terms of traps does not hold in this regime. Instead,
we propose that here activation is principally driven by entropic instead of energetic
barriers.  In particular, we illustrate that activation is not controlled by the 
hopping of high energetic barriers, and should more properly be interpreted as the entropic 
selection of nearly barrierless but rare pathways connecting metabasins on the landscape.
\end{abstract}

\maketitle

\section{Introduction}

The dynamics of supercooled liquids are slow, but not nearly as slow as the behavior described by mean field (MF) theory~\cite{cavagna:09,altieri:20}, which instead predicts the existence of extensive energy barriers between metastable states. The accepted reason for this difference is that processes not described by MF theory start dominating at low enough temperatures. The process generally implicated as most crucial is activated dynamics~\cite{arceri:20}.

In the broadest and least informative sense, activated dynamics can be any type of rare dynamical process that takes place over exponentially long time scales. These processes usually involve overcoming barriers in the free-energy landscape, which is more rugged as the temperature is lowered.
Given, however, the difficulty to access the free-energy landscape, it is commonly assumed that that at low temperature one can neglect the contribution of the entropy to the free energy, and focus on the potential energy~\cite{goldstein:69}. This is supported by evidence that below the onset temperature the dynamics seems to be dominated by the potential energy~\cite{sastry:98,schroder:00}.
Therefore, in the common view, activation is pictured to occur via the hopping of potential energy barriers. Namely, the system is stuck for long times in a potential energy well (a \emph{trap}), where it is confined by energy barriers that can be overcome only by a rare thermal fluctuation~\cite{dzero:05}. The time $\tau$ spent in these traps grows exponentially with an energy barrier $\Delta E$ and the inverse temperature $\beta$,\footnote{
Throughout the paper, we set the Boltzmann constant to unity, $k_\mathrm{B}=1$, 
so the temperature has the dimensions of an energy.}
as described by the Arrhenius law, $\tau\sim\exp(\beta\Delta E)$~\cite{arrhenius:1889}.

A simple toy model that has played an important role in the understanding of activated dynamics 
in a glassy energy landscape is the 
Trap Model (TM)~\cite{dyre:87,bouchaud:92}. In the TM, the phase
space is a fully-connected graph with each configuration assigned a random energy $E$. Transitions
from one configuration to another follow an Arrhenius law with energy barrier $\Delta E=\Eth-E$, and with
$\Eth=0$. For the usual case of an exponential distribution of trap energies, this simple model exhibits weak ergodicity breaking,  \emph{i.e.} the phase 
space is not fractured but cannot be fully sampled in finite times. In addition, this model provides a
series of non-trivial predictions regarding trapping times and autocorrelation functions which
can be used to guide the interpretation of activated dynamics in models of glasses, both in MF and in low spatial dimensions~\cite{bouchaud:92,monthus:96,denny:03}.

In MF, it has been shown that the behavior of the TM is quantitatively recovered in the 
Random Energy Model (REM)~\cite{gayrard:18,baityjesi:18} and similar 
models~\cite{benarous:08,gayrard:16,baityjesi:18c}. This does not appear to hold 
in the discrete $p$-spin model~\cite{stariolo:19,stariolo:20}, despite the fact that the 
$p$-spin model has a well-defined threshold energy which can be made to coincide with the condition $\Eth=0$.
In simulations of realistic 3D glass formers, there is a general consensus that activated processes are 
present, both below the dynamical transition 
temperature $T_c$~\cite{angelani:00,broderix:00,schroder:00}
and above it~\cite{denny:03,heuer:08}. 
However, in 3D some key ingredients of the TM do not hold.  For example, there does not seem to be a fixed threshold energy $\Eth$ that must be reached for a transition to occur~\cite{doliwa:03d}. Understanding the nature of the activated processes taking place above the Mode-Coupling temperature $T_c$ and the degree to which other aspects of the TM hold for realistic supercooled liquids will be our focus in this work.

In order to define traps in liquid-state simulations, it is advantageous to quench 
the system at every time step and study the local minima on the energy landscape (the 
\emph{inherent structures}, ISs) along the trajectory~\cite{sastry:98}.
It has been noticed that the ISs organize in superstructures, commonly called 
metabasins (MBs), and it has been argued that even above the mode-coupling temperature $T_{c}$ 
(but below the onset temperature $\To$ where supercooling starts), that barrier hopping between MBs drives 
the glassy dynamics, both in numerical~\cite{denny:03,doliwa:03,doliwa:03c,doliwa:03d}
and experimental systems~\cite{fris:11}.  This is somewhat surprising given that in the canonical Random First-Order Theory (RFOT), $T_{c}$ should mark the temperature below which barrier activation becomes important~\cite{rizzo:20}. An interpretive reconciliation between the observed MB dynamics and RFOT will be put forward at the end of this paper.

Despite the fact that visual inspection of MB dynamics seems to strongly suggest activated dynamics, some puzzles remain.
In fact, even through the lens of MBs, dynamical interpretations in terms of the TM have never been quantitatively 
satisfactory.  This fact has led both to the creation of more complicated trap models which 
incorporate the concept of MB in their definition~\cite{heuer:05},
or to a rejection of a landscape-based description of the dynamics via the invocation of kinetically constrained models that involve no static energy landscape yet predict more accurately observables such as the trapping time 
distribution~\cite{berthier:03b,berthier:05c}.

Here, we attempt to unify these viewpoints and solve the puzzles presented above {for $T>T_{c}$} through the analysis of simulations of a paradigmatic model of a glassformer. We show that while the underlying landscape directly relates 
to the glassy dynamics, the common concept of traps needs to be revised. Most importantly, instead of the system being
trapped between high energy barriers, it is more accurate to regard the system as dominated
by basins of attraction of the underlying landscape between which dynamics are slowed by the
search for increasingly rare directions in phase space. Some of our discussion will be based on new interpretations of old metrics, while some of our conclusions are prompted by completely new analysis.  

Our paper is organized as follows. In Sec.~\ref{sec:model} we briefly describe the model and the simulation protocols. In Sec.~\ref{sec:landscape} we treat the relationship between the landscape and dynamics, and in Sec.~\ref{sec:entropic} we show evidence that strongly implicates effects as the driver of activated dynamics. Finally, we discuss our results in Sec.~\ref{sec:discussion}.

\section{Model and Simulations}\label{sec:model}
We simulate a Kob-Andersen mixture~\cite{kob:94} of $N=65$ particles~\cite{doliwa:03b} of density $\rho=1.2$, at temperatures $T=5.0,2.0,1.0,0.8,0.7,0.6,0.49,0.46$.\footnote{
To ensure thermalization, we calculated the time scale $\hat\tau$ that it takes for the $F(\mathbf{k},t)$ to decay to values close to zero (this time scale is at least one order of magnitude larger than the characteristic $\alpha$-relaxation time scale $\tau_{\alpha}$). We then systematically simulated systems for $24\hat\tau$ in an NVT ensemble run with a Nos\'e-Hoover thermostat. Next we ran independent trajectories of length $3\hat\tau$ which were used to calculate the self-intermediate scattering function. Our thermalization criterion is that average self-intermediate scattering functions must be compatible within statistical fluctuations.} 
The cutoff radius is $r_\mathrm{cut}=L/2$, with $L=3.78364777565$, the linear size of the box consistent with the system density and number of particles, and the potential is shifted to be smooth at $r_\mathrm{cut}$.
Molecular dynamics simulations are run with a Nos\'e-Hoover thermostat, using the hoomd-Blue GPU package~\cite{anderson:08,glaser:15}, with a time step $dt=0.0025$~\cite{code:structuralglass}.

\begin{figure}[tb]
 \includegraphics[width=\columnwidth]{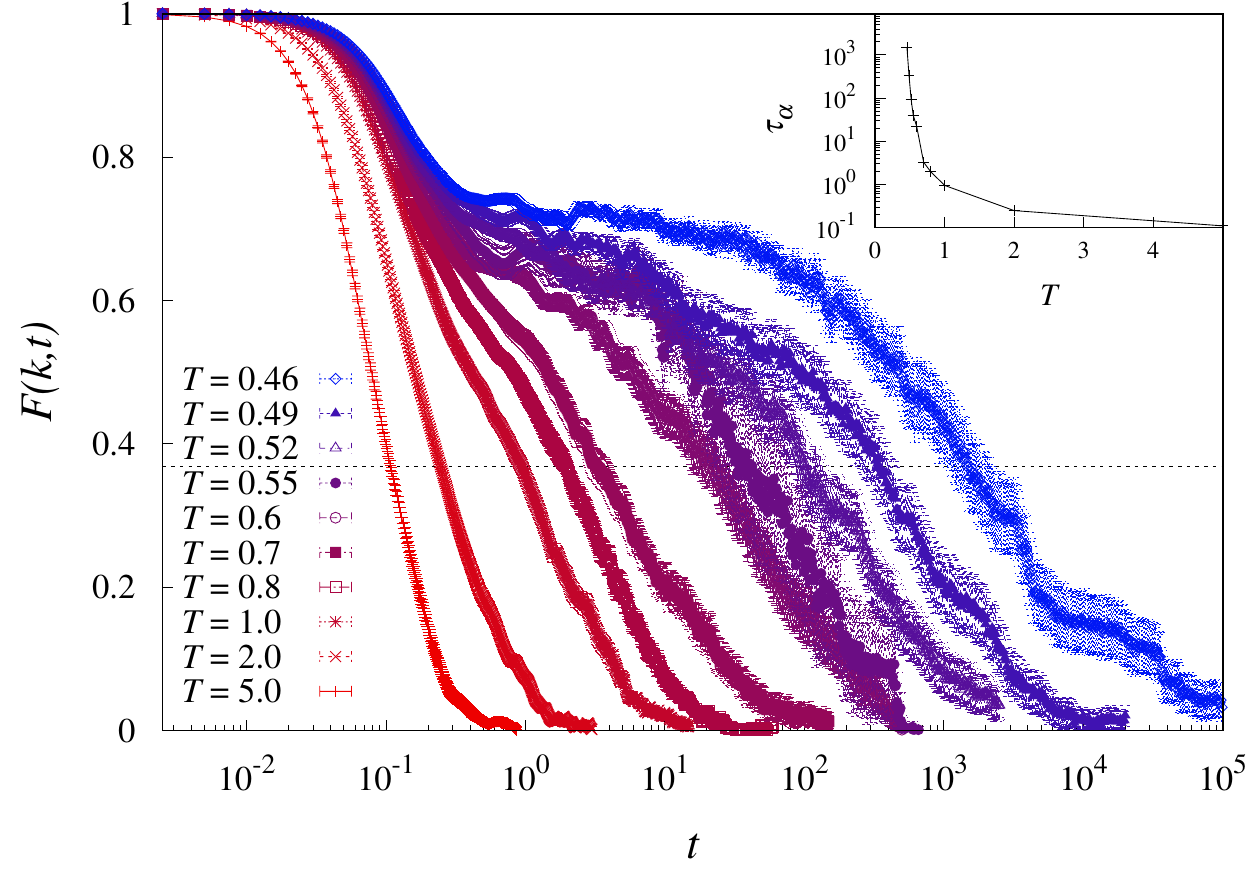} 
 \caption{$F(\mathbf{k},t)$ for all simulated temperatures, averaged over 10 different initial conditions. The dashed horizontal line indicates the value $1/e$. Inset: the autocorrelation time $\tau_\alpha(T)$.}
 \label{fig:fkt}
\end{figure}

To calculate the autocorrelation time $\tau_\alpha$, we measure the self-intermediate scattering function, $F(\mathbf{k},t)=\frac1N\sum_{i=1}^Ne^{i\mathbf{k}\cdot(\mathbf{r}_i(t)-\mathbf{r}_i(0))}$, where $\mathbf{r}_i(t)$ is the position of particle $i$ at time $t$, and $|\mathbf{k}|=\frac{2\pi}L|(1,3,4)|\simeq8.467$. The time $\tau_\alpha(T)$ is defined as the first time at which the average $F(\mathbf{k},t)$ crosses the value $1/e$ (Fig.~\ref{fig:fkt}).

For each $T$, we generate 10 independent initial conditions, and for each we run 10-20 NVT trajectories that are $2000\,\tau_\alpha$ long.
For each of these trajectories, we also calculate the \emph{inherent trajectory}, defined as the succession of the ISs related to each configuration in the trajectory (details on the inherent trajectories and MBs are given in App.~\ref{app:mb}).

\section{The influence of the landscape}\label{sec:landscape}
\subsection{Landscape signature of the onset of glassiness}
\begin{figure}[tb]
 \includegraphics[width=\columnwidth, trim=0 30 20 50]{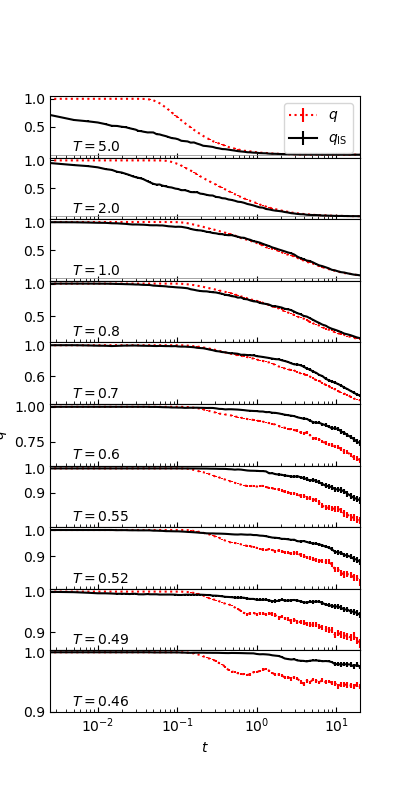}
 \caption{Overlap as a function of time for different temperatures. The red curves represent the overlap $q$ between the initial configuration and the configuration at time $t$. The black curves are the overlap $q_\mathrm{IS}$ between the initial configuration and the IS at time $t$. Note that close to the onset temperature $T_{o} \sim 0.8-1$, $q_\mathrm{IS}$ is closest to $q$.}
 \label{fig:qtraj}
\end{figure}

One expects that the potential energy landscape plays a progressively more dominant role as $T$ is decreased.  Thus ISs can provide useful information on how the landscape drives the glassy behavior. To illustrate this, we define the overlap
\begin{equation}\label{eq:q}
 q(t_0,t)=\frac1N\sum_{i=1}^N \Theta\left(|\mathbf{r}_i({t_0})-\mathbf{r}_i(t)|-a\right)
\end{equation}
between times $t_0$ and $t$.~\footnote{
We set $a=0.3$, as it is a common choice used in literature~\cite{karmakar:09}.
} 
We can see a clean signature of the onset of glassiness if we compare the evolution of the overlap between thermal configurations, $q(0,t)$, with those from the associated inherent trajectories, $\qis(0,t)$ (Fig.~\ref{fig:qtraj}).  The approach taken here is inspired by, and very similar to, that presented in~\cite{schroder:00}.  However by focusing on the overlap and not the IS intermediate scattering function, one can cleanly estimate the location of the onset temperature $T_{o}$, as seen in Fig.~\ref{fig:qtraj}.  The interpretation of this behavior is discussed below.

At every temperature, the overlap $q(0,t)$ is equal to 1 at short times, since no particles have moved by more than $a$, and at later times exhibits a clear crossover towards lower values (with a plateau appearing at lower temperatures). 
When we look at the short-time dynamics of $\qis(0,t)$, we need to distinguish between high and low temperatures. At high temperatures, $\qis(0,t)<1~~\forall t$, whereas at low temperatures it remains close to 1 for progressively longer times. This is expected in a landscape description of the dynamics.  In particular, at high temperatures the dynamics is independent of the landscape, and at every time step the system finds itself in a different basin of attraction. Therefore, minimizing the energy {\em increases} the distance between two initially nearby configurations.  On the other hand, at low temperatures, the dynamics are driven by the underlying landscape, so minimizing the energy leads initially far-away configurations \emph{towards} the same attractor state.\footnote
{
A similar phenomenology, that has enabled the identification of the onset temperature via the overlap, has recently been observed in diverse systems, ranging from the mixed $p$-spin model~\cite{folena:20} to the 3D Heisenberg spin glass~\cite{baityjesi:19b}.
}
As a consequence, the difference between $q(0,1)$ and $\qis(0,1)$ 
can be used to detect the onset of glassiness,
and shows that glassiness begins at $\To\simeq1$, 
when the landscape starts playing a role. 
Of course, we can also locate the onset of glassiness without the use of ISs by finding the temperature where
the self-intermediate scattering function begins to develop
a plateau (Fig.~\ref{fig:fkt}).


\subsection{Metabasins}
\begin{figure}[tb]
\includegraphics[width=\columnwidth]{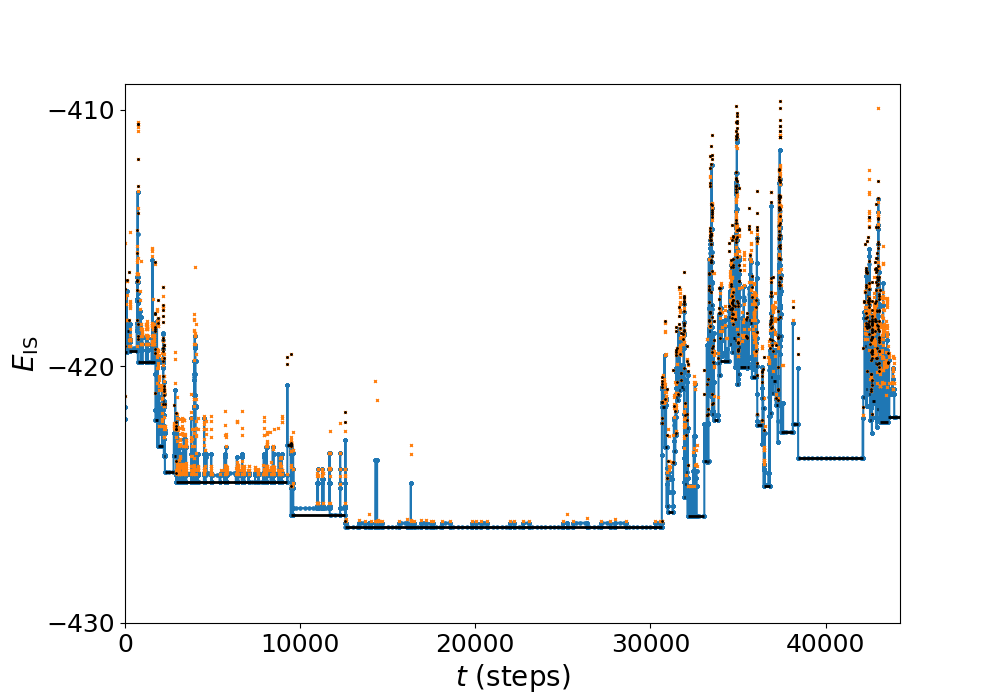}
 \caption{Inherent structure trajectory for $T=0.6$. The blue points are the ISs, the orange points are the ridges (\emph{i.e.} the highest potential energy points along the separation between two subsequent ISs) between subsequent distinct ISs, and the black lines indicate the MBs, calculated with the procedure described at the beginning of App.~\ref{app:mb}.}
 \label{fig:MBtraj}
\end{figure}

A portion of a typical inherent trajectory is shown in Fig.~\ref{fig:MBtraj}.
As remarked in several previous works~\cite{heuer:08}, even though we are above the mode-coupling temperature $T_{c}$, the dynamical inherent trajectory reveals a remarkable structure, in which successions of ISs can be grouped into MBs. 
In Fig.~\ref{fig:MBtraj}, MBs are emphasized with black horizontal lines.

\begin{figure}[tb]
 \includegraphics[width=\columnwidth, trim=10 0 30 0]{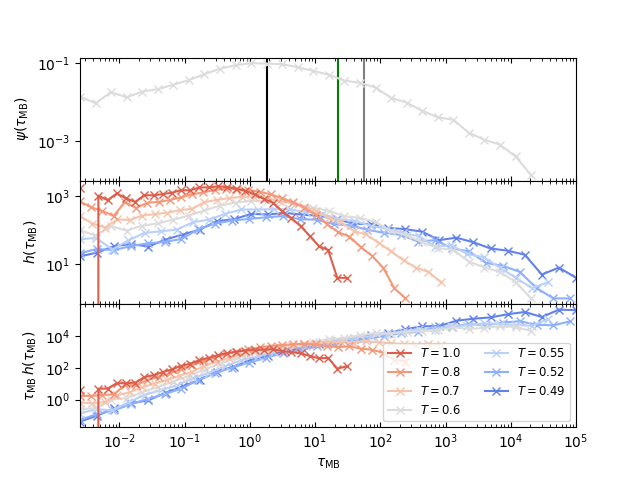}
 \caption{\textbf{Top}: $\psi(\tmb)$ for $T=0.6$. Vertical lines indicate $\tau_\mathrm{MB,1/2}$ (black), $\ta$ (green) and $\langle\tmb\rangle$ (grey). \textbf{Center}: Histogram $h(\tmb)$ of the MB times for all temperatures. \textbf{Bottom}: Histogram of the MB times weighted by the waiting time, $h(\tmb)\tmb$.}
 \label{fig:psitau}
\end{figure}

There is general agreement that MBs are the landscape structure that dominates the dynamics below $T_{o}$~\cite{heuer:08}. One can argue in favor of this viewpoint by calculating the time that the system remains in each MB,
 and comparing it to $\ta$. In Fig.~\ref{fig:psitau} we show the distribution of MB waiting times $\psi(\tmb)$ for $T=0.6$, highlighting that $\ta$ falls between the median ($\tmb^{1/2}$) and the average MB time scale ($\langle\tmb\rangle$).

\begin{figure}[tb]
\centering
\includegraphics[width=1.0\columnwidth]{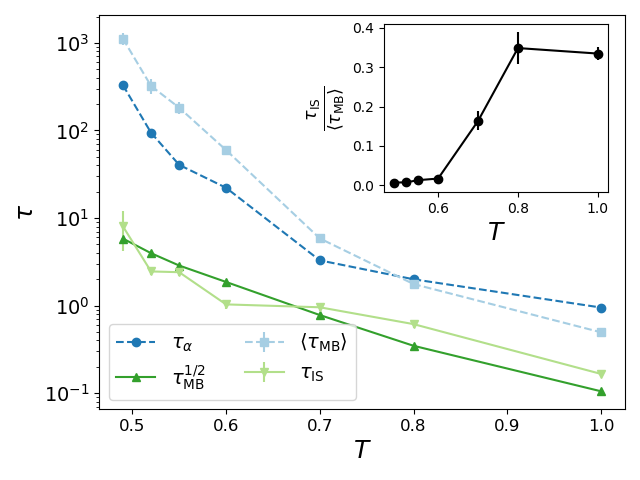}
\caption{Autocorrelation times $\ta$ (from $F(k,t)$, darker blue circles), $\langle\tmb\rangle$ (average time in MBs, lighter blue squares), $\tmb^{1/2}$ (median time in MBs, darker green upwards triangles) and $\tis$ (average time in ISs, ligher green downwards triangles), as a function of temperature. \textbf{Inset}: The ratio $\tis/\langle\tmb\rangle$ indicates the onset of glassiness.}
\label{fig:taus}
\end{figure}

The correspondence between MBs and $\ta$ is verified both below and above the onset temperature $\To$.
In Fig.~\ref{fig:taus}, we show that $\ta$ and $\langle\tmb\rangle$ are similar over the full temperature range, suggesting that $\ta$ is directly correlated with the MB dynamics. 
We also show the median MB time, $\tmb^{1/2}$, and the average time spent in the ISs, $\tis$. The latter two time scales also have mutually similar values at all temperatures, and once $T<0.8 \sim T_{o}$, they both become consistently smaller than the former two time scales (Fig.~\ref{fig:taus}--inset).

The fact that $\langle\tmb\rangle$ and $\ta$ grow much faster than $\tmb^{1/2}$ does with lowering temperature indicates that the typical\footnote{ 
The word \emph{typical} is usually used for the argmax of the distribution. In our case the peak and median of the distribution almost coincide, so we can also use the word \emph{typical} here.
} 
behavior is not strongly influenced by temperature, and that the strong slowing of dynamics in supercooled liquids can be be attributed to a few very long-lived MBs (the tail of $\psi(\tmb)$).  This behavior becomes more prominent as $T$ decreases. 

Furthermore, the wide tails of $\psi(\tmb)$ (Fig.~\ref{fig:psitau}) extend several orders of magnitude beyond $\ta$. 
This means that the MB dynamics are dominated by rare structural configurations which last hundreds of times longer than $\ta$.
As a consequence, the average $\ta$ is not a good indicator of the time scales for which the system can remain blocked. This is in line with recent work demonstrating that the dynamics of supercooled liquids is dominated by broad distributions (including of $\ta$), characterized in a somewhat distinct manner~\cite{berthier:20}. 

 To make this point even clearer, in Fig.~\ref{fig:psitau} we plot the quantity $\psi(\tmb)\tmb$,\footnote{We show the histogram $h(\tmb)$ instead of the density $\psi(\tmb)$, because the whole integral cannot be calculated for all temperatures.}
which represents the weight that each value of $\tmb$ has on the average $\tmb$. The peak
of these quantities is much larger than $\ta$, indicating that even though the typical metabasin reflects $\ta$, the average is dominated by times that are much larger.
This also indicates that the estimators of the average $\tmb$ are biased due to an insufficiently long simulation time (we can speculate that an unbiased estimator at $T=0.49$ would require sampling basins that last $\sim 10^{12}$ LJ units).  An extreme version of the behavior manifested above can be found in the TM, where the system spends most of the time in the deepest trap and only a negligible time outside of it~\cite{bouchaud:92}.

\section{Entropic Effects}\label{sec:entropic}

\subsection{No Threshold}
The usual conception of a trap is that of a local minimum on the energy landscape that cannot be escaped unless a potential energy barrier is overcome.
In the TM and other MF models, all traps are escaped at the same energetic
height, called the threshold energy $\Eth$~\cite{bouchaud:92,rizzo:13,baityjesi:18}.
In the TM the threshold energy is purely based on the potential energy landscape and is thus independent 
of temperature.
If the concept of the threshold energy is applicable in realistic $3D$ supercooled liquids, then
the TM threshold energy would represent the energy of the ridge separating 
different metabasins, $\Er$ (the ridge points are the points that, along the trajectory, mark the separation of subsequent ISs or MBs in the potential energy landscape).\footnote{
We calculate the ridge energy $\Er$ through the procedure proposed in Ref.~\cite{doliwa:03c}, which consists of minimizing two adjacent configurations that lead to different ISs, until their distance becomes larger than 0.001. In order to obtain a higher estimate of the barrier, which takes into account thermal fluctuations around the transition path, we omit the final minimization of the gradient~\cite{code:structuralglass}.
}

\begin{figure}[t]
 \includegraphics[width=\columnwidth]{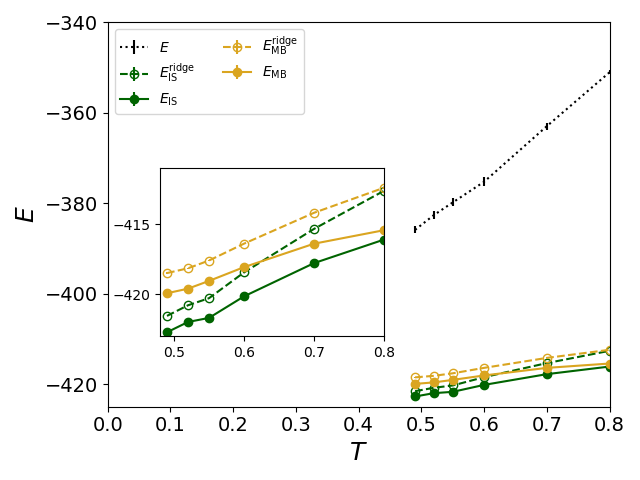}
 \caption{Thermal energy (black dots), IS energy (green), MB energy (yellow). Dashed lines are the ridges. The \textbf{inset} is a close-up. 
 Note that $\Eis$ is lower than $\Emb$ because the average over the ISs gives more weight to the ISs belonging to the deepest MBs, since those contain a larger number of ISs.}
 \label{fig:enerMB}
\end{figure}

At each temperature, for each simulated thermal trajectory we calculated the ridge between 
subsequent ISs and between subsequent MBs. This is shown in Fig.~\ref{fig:enerMB}, along with the average IS energy, $\Eis(T)$, and the
average MB energy, $\Emb(T)$. We notice that all four mentioned observables
depend on $T$. A dependence of $\Eis(T)$ on temperature is expected~\cite{sastry:98},
as is that of $\Emb(T)$~\cite{doliwa:03c}. However, we also note that the
ridge energies decrease with $T$, {at variance} with the standard TM.
This suggests that as the temperature is lowered, it becomes more convenient for the system to 
search for rare, lower energy pathways.
Therefore, there is no single energy
level that must be overcome in order to achieve barrier hopping, and the concept of $\Eth$ can not be used to describe this system, {at least in the temperature regime we focus on}. 
This observation is not new, and similar conclusions were obtained in 
Ref.~\cite{doliwa:03c} by looking at the relationship between $\Emb$ and the barrier height.\\

\subsection{No Traps}
In Fig.~\ref{fig:enerMB} we plot the total potential energy $E(T)$. A striking
feature is that $E(T)$ is significantly larger than the energy at the ridge between ISs, $\Er$, and at the ridge between MBs,
$\Er^\mathrm{MB}$.
Further, the difference $E(T)-\Er$ is significantly larger than $\Er^\mathrm{MB}-\Emb$.\footnote{ 
We realize that these same observations can be drawn from data found in previous works.
For 3D binary mixtures, we focus on Refs.~\cite{doliwa:03b,doliwa:03c}, which use 
models similar to ours. 
With $N=65$, at $T=0.6$, these authors find $E(T)\approx-230$ (Fig.~3b from~\cite{doliwa:03b}).
as well as IS and MB energies of $\Emb\approx\Eis\approx-295$ (Fig.~2a from from~\cite{doliwa:03b} 
and Fig.~3a from~\cite{doliwa:03c}). The energy of the barriers is 
$\Delta E=\Er^\mathrm{MB}-\Emb\simeq6.9$ (Figs.~12 (and 14) from~\cite{doliwa:03c}).
The ridge energy is thus $\Er^\mathrm{MB}=\Emb+\Delta E\approx-288\ll E(T)$.
A comparison between $\Eis(T)$ and $E(T)$ at $T=0.5$ is provided in the caption of Fig.~8 of Ref.~\cite{doliwa:03c}.
} 
In other words, the energy  of typical configurations is significantly larger than 
the energy separating neighboring MBs.
This means that, at least in the temperature range considered, 
we cannot think about activated dynamics as arising from the system being confined between high barriers 
which can be overcome via instantonic fluctuations of the energy. Instead, we argue that at these temperatures the dynamics is controlled by basins of attraction and the selection of rare, potentially low barrier pathways between them. 

{Since our systems are very small and
the coherence length is of the order of the linear system size $L$~\cite{yaida:16}, 
i}n order to satisfy the typical scenario of a system completely confined between energy barriers which can be overcome only through thermal activation, the system should be at a temperature $T$ such that $E(T)$ is smaller than $\Er^\mathrm{MB}(T)$.\footnote{
This is a necessary but not sufficient condition.
}
As Fig.~\ref{fig:enerMB} shows, in our system this can only happen at very low temperatures if the extrapolation from high temperatures are indicative of behavior below $T_{c}$.
These temperatures are significantly lower than that which can be probed directly by computer simulation. Hence, even though a TM paradigm might apply at low temperatures, it cannot hold in its simplest form in the range $T_{c} \leq T \leq \To$.

\begin{figure*}[tb]
\includegraphics[width=.9\textwidth]{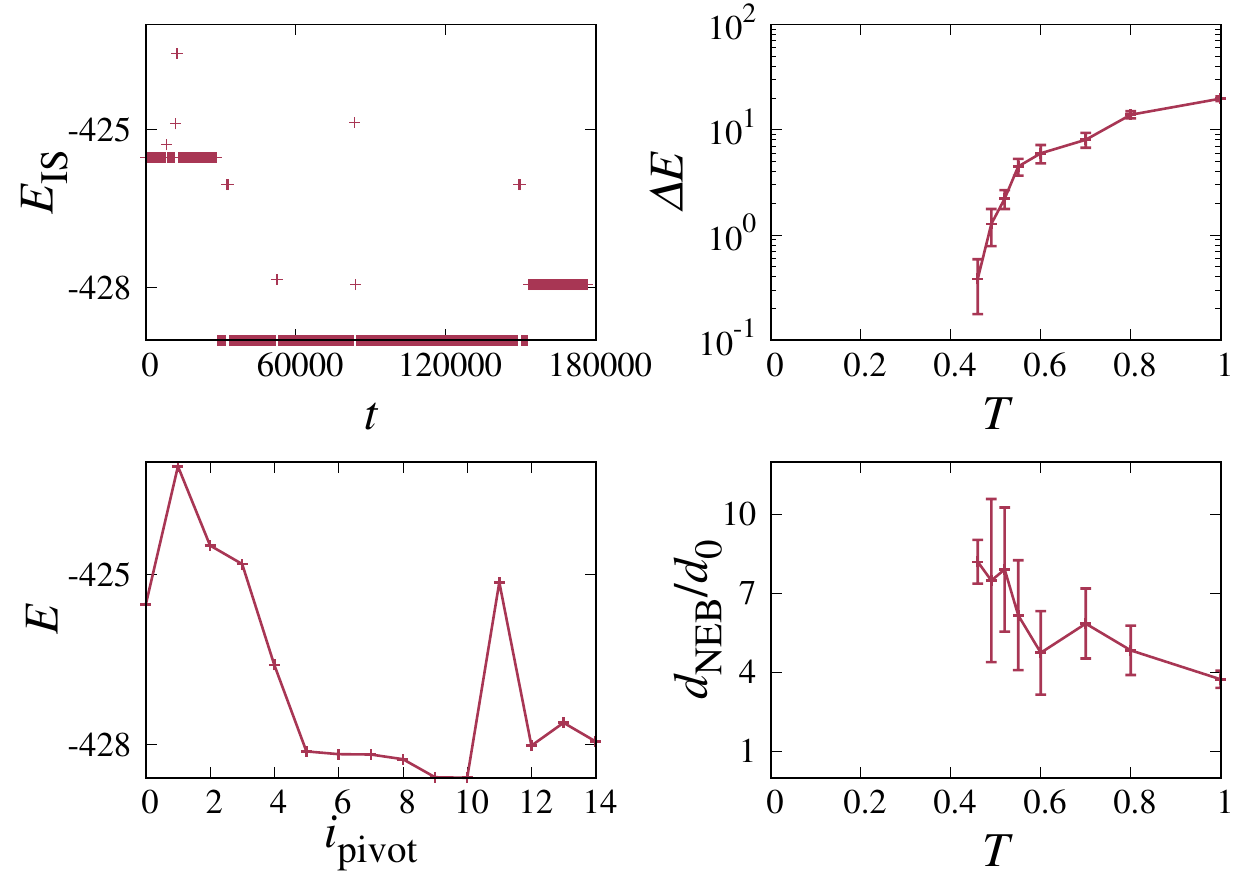}
  \caption{ \textbf{Top-left}: an IS trajectory at $T=0.6$. \textbf{Bottom-left}: 
  Nudged Elastic Band (NEB) interpolation of the minimum energy pathway for the same trajectory. The initial condition for the 
  NEB pathway is the thermal trajectory, as described in the main text. 
  \textbf{Top-right}: energy barrier of the NEB pathway at several temperatures, 
  averaged over 20 exit trajectories.
  \textbf{Bottom-right}: length $d_\mathrm{NEB}$ of the NEB pathway at several 
  temperatures, with 15 pivots, divided by the distance between 
  $\phi_\mathrm{ini}^\mathrm{IS}$ and $\phi_\mathrm{fin}^\mathrm{IS}$.}
  \label{fig:NEB}
\end{figure*}

\subsection{The Search for More Convenient Pathways}

To test the possibility that the MB dynamics are dominated by entropic effects, 
we run the following experiment. We first take a configuration $\phi_\mathrm{ini}^\mathrm{IS}$ at the bottom of a single deep 
MB. We then add thermal agitation to the configuration $\phi_\mathrm{ini}^\mathrm{IS}$, corresponding to a 
temperature $T$, and obtain $\phi_\mathrm{ini}=\phi(t=0)$. We next let the system 
evolve until ${t}_\mathrm{fin}=20\ta$, obtaining $\phi_\mathrm{fin}=\phi(t_\mathrm{fin})$. 
Finally, we minimize the energy again, obtaining $\phi_\mathrm{fin}^\mathrm{IS}$.
In Fig.~\ref{fig:NEB} we show an inherent trajectory obtained in this way.

We calculate the barrier between $\phi_\mathrm{ini}^\mathrm{IS}$ and 
$\phi_\mathrm{fin}^\mathrm{IS}$ using the climbing image Nudged Elastic Band (NEB) 
method~\cite{jonsson:98,draxler:18}.\footnote{The climbing image method can be summarized as follows: After a number of NEB steps, the highest-energy pivot aims at increasing its energy instead of decreasing, in order to better estimate the barrier height at its peak. {The barriers estimated through NEB and ridge method are consistent when looking at transitions between neighboring ISs, but the NEB is more suited than the ridge method
for the zero-temperature MB transitions that we consider in this section.}} 
In the standard formulation of the NEB, 
one creates $\np$ ordered replicas of the system, called \emph{pivots}, which interpolate
between $\phi_\mathrm{ini}^\mathrm{IS}$ and $\phi_\mathrm{fin}^\mathrm{IS}$. 
The NEB pathway is obtained by optimizing a system where each pivot feels its own potential
energy as well as an attraction to its two closest neighboring pivots.\footnote{
We use the climbing image variant of the NEB, which attracts the 
highest-energy pivot towards higher energies, 
in order to find the highest point in the trajectory.
} 
The resulting path
depends on the number and the initial configurations of the pivots.
A common choice for the initial configuration of the pivots is a linear interpolation
between $\phi_\mathrm{ini}^\mathrm{IS}$ and $\phi_\mathrm{fin}^\mathrm{IS}$. 
Here, since we are interested in calculating the barrier that is surmounted during the
dynamics, we use the thermal trajectory for the initial configuration of the pivots.
We set the first pivot to $\phi_\mathrm{ini}^\mathrm{IS}$ and the last one to 
$\phi_\mathrm{fin}^\mathrm{IS}$. The starting configurations of the intermediate 
pivots are equally spaced configurations along the thermal trajectory between 
$\phi_\mathrm{ini}$ and $\phi_\mathrm{fin}$.
We show an example of the final NEB profile using these 
starting pivots in Fig.~\ref{fig:NEB}. 

In Fig.~\ref{fig:NEB} we see that the average NEB barrier (the difference between the 
highest energy of the NEB and $\phi_\mathrm{ini}^\mathrm{IS}$) decreases in size as $T$ is decreased. This is in agreement with the observation that $\Er(T)$ decreases
with cooling (Fig.~\ref{fig:enerMB}), but is not in agreement with the conclusions of Ref.~\cite{doliwa:03c} where it is found that the barrier height, which is approximately the average over $T$ of the one we find, is independent of temperature.  We return to this discrepancy before concluding.  

Our results suggest that low-energy pathways exist, but they are rare and it takes a very long time to find them. If the thermal agitation is high compared to $\Delta E$, the barriers are jumped before a low-energy pathway is found. If the typical $\frac{\Delta E}{T}$ is high, barrier hopping becomes unlikely, and the system has the time to search for the low-energy pathways.

In Fig.~\ref{fig:NEB} we also show the total length $d_\mathrm{NEB}$ 
of the NEB pathway, divided by the distance $d_0$ between 
$\phi_\mathrm{ini}^\mathrm{IS}$ and $\phi_\mathrm{fin}^\mathrm{IS}$. 
The ratio $d_\mathrm{NEB}/d_0$ passes from 4 at $T=1$, to around 8 at $T=0.46$, indicating 
that the system is wandering progressively in a more tortuous manner as temperature is lowered in order to find more convenient pathways.

We have also tried to perform the same type of analysis with NEB pathways starting from a linear interpolation
between $\phi_\mathrm{ini}^\mathrm{IS}$ and $\phi_\mathrm{fin}^\mathrm{IS}$, in order
to show that in this case the barrier is independent of $T$. However, given the long 
distance and complicated landscape between the two configurations, the linear 
interpolation passes through highly unphysical configurations with extremely high
energies ($\sim+10^6$), and we were not able to reach convergence for the energy pathways with a reasonable number of pivots.\\

\subsection{Metabasin Dynamics and Localization}\label{sec:hopping}
\begin{figure}[tb]
 \includegraphics[width=\columnwidth, trim=0 0 30 30]{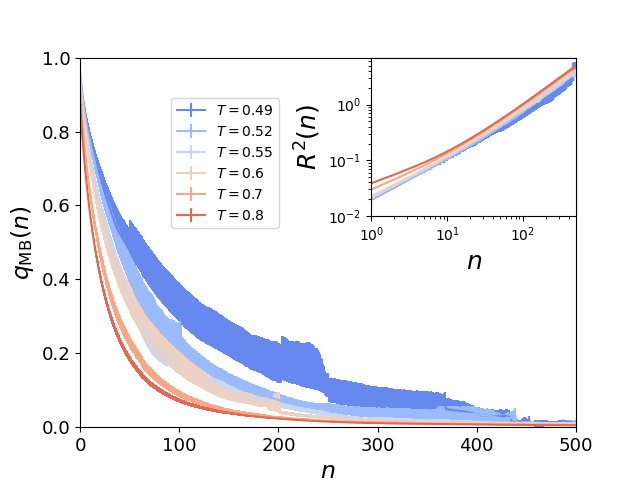}
 \caption{Metabasin overlap $\qmb(n)$ as a function of the number of MB transitions $n$, for different temperatures. {In the inset we show the mean squared displacement $R^2(n)$ after $n$ MB transitions, for the same temperatures. While $R^2(n)$ is roughly independent of $T$, $\qmb$ exhibits a clear dependence. Both quantities are normalized by a factor $N$.}}
 \label{fig:qMB}
\end{figure}

We now turn to the relationship between subsequent MBs in a MB trajectory.
In Fig.~\ref{fig:qMB} we report the overlap between MBs,
$\qmb(n)$,\footnote{
We calculated $\qmb$ using Eq.~\eqref{eq:q} between two MBs A and B both as the average overlap between all the ISs in A with all the ISs in B, and as the overlap between the deepest configuration of each MB. We did not notice significant differences between the two procedures. The data shown are for the latter procedure.
}
as a function of the number of MB transitions $n$. 
As expected, $\qmb(n)$ decreases monotonically, and the system 
decorrelates progressively over time. However, the curves have a clear dependence on temperature. These curves do not depend on the amount of time 
spent in each MB, so the longer decorrelation time
at lower $T$ implies that the nature of metabasin transitions takes on a different character 
as $T$ is lowered.
At first sight, this result seems to contradict 
previous observations that the squared displacement $R^2(n)$ after a fixed number of MB
transitions has a weak dependence on $T$~\cite{doliwa:03d}, as shown in the inset of Fig.~\ref{fig:qMB}.
However, notice that $\qmb(n)$ and $R^2(n)$ are different indicators of the system's 
particle motion, and they are influenced by the kind of rearrangements that take place in a different manner. 
For example, a large displacement of a single particle leads to a large 
variation in $R^2(n)$, but only a $\sim1/N$ variation in $\qmb$. The discrepancy 
between $\qmb(n)$ and $R^2(n)$ is thus an indication of the {\em localization} of the particle 
rearrangements along the dynamics, which must be considered with special care.
{In fact, in order to decrease the overlap, several particles need to move, so if the decay becomes slower, this indicates that fewer particles are moving by a significant amount. If $R^2(n)$ does not change with lowering $T$, this is suggesting a crossover from a regime in which all particles contribute equally to $R^2(n)$, to one where the displacement is dominated by few particles.}
We reserve the analysis of the localization of particle motion during MB transitions for future.
Regardless, when viewed from the perspective of the overlap, we see that the assumption of the TM
that there should be near-instantaneous temporal decorrelation of MB does not hold in our system.

\section{Discussion}\label{sec:discussion}
In this work we have revisited the relationship between the landscape and activated processes at 
temperatures higher than the mode-coupling transition temperature, $T \geq T_{c}$. 
We began by exposing some signatures of the onset of glassiness that stem 
from the potential energy landscape.  In particular:
(i) on time scales of order $t\sim1$ LJ step, the overlap is $q(0,t)>\qis(0,t)$ in the high temperature
liquid phase, but the relation is inverted at lower temperature, once the dynamical slowing down is 
driven by the underlying landscape.
(ii) The ratio between the typical (median) MB time, $\tmb^{1/2}$, and the 
average MB time, $\langle\tmb\rangle$, drops at the onset temperature, heralding the initial growth of wide tails in $\psi(\tmb)$. 

The relationship between the landscape and dynamics is highlighted by the fact that
$\langle\tmb\rangle \approx\ta$ at all temperatures. As $T$ decreases, $\langle\tmb\rangle$ grows 
faster than  
$\tmb^{1/2} (\approx\tis)$. This different rate of growth indicates that even 
though typical configurations relax more slowly upon cooling, the strong dynamical
arrest of glasses is driven by the tails of the distribution $\psi(\tmb)$. Furthermore, even $\ta$
falls short of properly describing the slowing down of supercooled liquids, since the tails of $\tmb$ are
so wide that one encounters with finite probability MBs that live several orders
of magnitude longer than $\ta$.

We then analyzed the applicability of a picture based on energetic barrier hopping. 
We find that the energy of the ridges between MBs, $\Er(T)$, decreases with 
temperature and, most strikingly, the potential energy of the system satisfies
$E(T)>\Er(T)$. In other words, the system is always {\it above} the ridge
between MBs. 
However, as others have previously remarked, we find that the
landscape does play a role. The emerging picture is that the MB structure
in the IS energy time series, $\Eis(t)$,
does not arise from confinement between energy barriers, but rather
from groups of IS that lie in the basin of attraction of MBs with limited escape routes in a high dimensional space. As $T$ decreases, the system eventually finds lower-energy pathways from one MB to another. Thus,  MB transitions do not occur at the lowest possible energy and instead as $T$ is lowered it becomes more convenient for the system to search for alternative pathways. This means that there \emph{is} an entropic cost associated with 
MB transitions even though the barrier is much lower than the typical energy.
This observation is consistent with previous work which argued that, above $T_c$, any two configurations can be connected by a barrierless geodesic, and that the slowing down should be attributed to the geodesics becoming more tortuous~\cite{wang:07,wang:07b,nguyen:12}. 
However, as our work suggests, even if a barrierless geodesic exists, it might be too hard for the system to find such a path. {Actually, the more likely situation is that there exists an exponential number of energetic barriers, as found in mean-field models \cite{ros2019complexity}, and the competition between their height and their entropy plays a key role in activated dynamics. In the regime of temperature that we are considering, the {competition between different transition states}
largely dominates, but it diminishes at lower temperature, hence the system is pushed to find and cross lower energetic barriers.}
Consequently, we can think that the effective dimensionality of the system decreases upon cooling, which is consistent with a previous picture based on TM-like aging functions in
supercooled liquids~\cite{fabricius:04}.

Even though our core numerical findings are not different from those found in previous seminal studies~\cite{doliwa:03,doliwa:03b,doliwa:03c,doliwa:03d,heuer:08}, our interpretation is different. 
In addition, we find differences in the temperature dependence of extracted barrier heights, which appear to be independent of $T$ in Ref.~\cite{doliwa:03c} but decrease with decreasing $T$ here. This distinction may be due to the different ways barriers are extracted in Ref.~\cite{doliwa:03c} versus in this work. We note that while in Ref.~\cite{doliwa:03c} the barrier height is independent of temperature, it does correlate with $\Emb$, which is dependent on temperature. This is somewhat surprising since it implies that $\frac{d\Emb}{dT}=\frac{d\Er}{dT}$.  These two quantities are not expected to be related in any known landscape-based model of glassy systems.

We also find that the overlap {$\qmb(n)$} between subsequent MBs decorrelates more slowly as $T$ is lowered, which at first sight appears to be in contradiction with the observation that the squared distance $R^2(n)$ between subsequent MBs is almost independent of $T$~\cite{doliwa:03d}.\footnote{{
When viewed through the lens of the the quantity $R^2(n)$, MB dynamics appear to resemble a random walk~\cite{doliwa:03d,rubner:08}, but not if we focus on $\qmb(n)$.  It should be noted that this apparent random walk is not a renewal process, as would be required by the TM. 
}}
However, these two observables should only behave in the same manner in the case of fully delocalized displacements.
The fact that, upon cooling, $\qmb(n)$ decays progressively more slowly while $R^2(n)$ remains constant, {seems to indicate} that MB transitions become {less} collective, with smaller per-particle displacements.
This is consistent with past observations showing more localized motion approaching $\Td$.
In fact, transitions are string-like in this regime~\cite{donati:98}, and a localization transition takes place at $\Td$~\cite{coslovich:19}.

The above observations suggest that one should regard a supercooled liquid as confined by large barriers in most, but not all, directions, and that the slowdown of 
dynamics is mainly driven by entropy. 
The residual small barriers that need to be 
overcome can likely be explained by localization effects, which should be taken into
account in a MB description of the dynamics.  

We can use these ideas to analyze an argument that has been advanced against the importance of the growth of a static length scale (and thus a potential energy landscape picture) in the slowdown of supercooled liquids~\cite{wyart:17}. This argument is based on the observation that the introduction of swap dynamics~\cite{grigera:01,ninarello:17} suppresses the glass transition temperature without altering the free energy landscape. This argument could be tested by looking at the inherent trajectory of a swap simulation. 
A general expectation would be that swap dynamics removes energy barriers, so the MB structure of $\Eis(t)$ would be disrupted. However, we have argued that the slow dynamics are driven by entropic rather than energetic effects, so the MB structure may  survive. If it does not, the landscape-based indicators of $\To$ described here could be useful in telling us if one can attribute this disruption merely to a shift in $\To$~\cite{berthier2019can,ikeda:17} caused by the addition of a larger number of favorable pathways to relaxation.

The view that the activated dynamics is driven by entropic rather than energetic barriers
is further supported by a recent numerical investigation on the discrete 
3-spin model~\cite{stariolo:20}, 
where it was demonstrated that a MB structure is visible in the inherent
trajectory, and that observables such as trapping time distributions are not 
predicted as well by the TM as they are by the
Step Model (SM). 
The SM is a simple toy model with a single energy minimum, where the exponentially slow dynamics
are purely driven by the scarcity of low energy configurations~\cite{barrat:95}. 
We also remark that in some temperature regimes entropy-driven activation can indeed resemble
energy-driven activation~\cite{bertin:03,cammarota:15,cammarota2018numerical,carbone:20}, and can be translated into a competition
between energy and high dimensionality~\cite{carbone:20, tapias:20}, whereby an aging system progressively passes from entropy- to energy-driven dynamics~\cite{tapias:20}.

In conclusion, we argue that in the temperature range $T_{c}<T<\To$, the potential energy landscape indeed plays an increasingly important role in driving the dynamics of supercooled liquids, but the nature of the landscape picture based on MBs needs to be revised {in light of the following observations:
\begin{itemize}
    \item The potential energy is always higher than the energy of the barriers between metabasins.
    \item The ridge energy of MB transitions depends on the temperature, indicating that the system surmounts barriers only when it is more convenient than searching for rare   pathways with a low barrier.
    \item The MB energy barriers are lower in cooler systems, but the pathways are longer.
    \item The overlap between subsequent metabasins decays more slowly as temperature is decreased.
\end{itemize}
}
These findings suggest that, in this regime, the state of the system does not appear to be confined between high energy barriers, but rather inside the basin of attraction of large MBs, where it remains for times that are often much larger than $\ta$. Here, energy barriers are small, and the dynamical slowdown is therefore attributed to the scarcity of energetically favorable pathways between MBs. This interpretation helps rationalize why the thermal and inherent trajectories appear to be qualitatively so different, and why the MB structure only arises in the latter, a fact in contrast with all models known to exhibit TM-like dynamics.

{For $T<T_c$, one could argue that the situation stays similar.}
{However, at $T_c$ the nature of the landscape probed during the dynamics seems to change and a geometric/localization transition takes place~\cite{angelani:00,broderix:00,fabricius:02,coslovich:19}. Random First Order Transition theory \cite{kirkpatrick:89} suggests that at these lower temperatures free-energy barriers become mainly energy barriers, i.e. their entropy plays a sub-dominant role. In this regime, activation would then correspond to barrier hopping, and concomitantly a picture based on classical TM-like energetic physics may apply~\cite{schroder:00}.}


\begin{acknowledgments}
We thank F. Landes and S. Sastry for valuable discussions.
This work was funded by the Simons Foundation through
the collaboration “Cracking the Glass Problem” (M.B.J. and D.R.R. funded by award No. 454951 and G.B. funded by award No. 454935). This work benefited from access to the University of Oregon high performance computer, Talapas.
\end{acknowledgments}

\appendix

\section{Numerical Calculation of the Metabasins}\label{app:mb}
To minimize the energy we used the FIRE algorithm.\footnote{With parameters $dt$=0.0025, $\alpha_\mathrm{start}$=0.99, $f_\mathrm{tol}=10^{-5}$, $E_\mathrm{tol}=10^{-10}$, $w_\mathrm{tol}=10^{-5}$.}\\
To find the inherent trajectory, instead of minimizing the energy at every single time step, we used a bisection procedure essentially identical to that described in Ref.~\cite{doliwa:03c}. We separated the trajectory in intervals of $10^5$ MD steps, and calculated the IS at the beginning and at the end of the interval~\cite{code:structuralglass}. If the configuration does not change,\footnote{
Analogously to Ref.~\cite{doliwa:03c}, we say that two ISs $\xi$ and $\xi'$ are the same if their energies $E$ and $E'$ are the same, $|E-E'|<\epsilon$, with a resolution $\epsilon=10^{-5}$.
}, we postulate that all the configurations in between are the same.
If the configuration does change, we repeat the procedure for the two resulting intervals. This procedure is iterated for the increasingly smaller intervals, and is stopped when either the intervals have the same energy at the beginning and at the end, or they are one step long.\\
To identify the MBs, we calculate the minimum energy along the whole trajectory. All the configurations between the first and the last occurrence of this configuration are said to belong to the same MB. We then calculate the minimum energy along each of the remaining intervals, defining the MB time in the same way. The procedure is stopped when there are no more ISs with negative energy, or more than 3000 MBs are found (which only happens for $T\geq1$).

\paragraph{On the definition of MBs}

\begin{figure}[bth]
\includegraphics[width=\columnwidth]{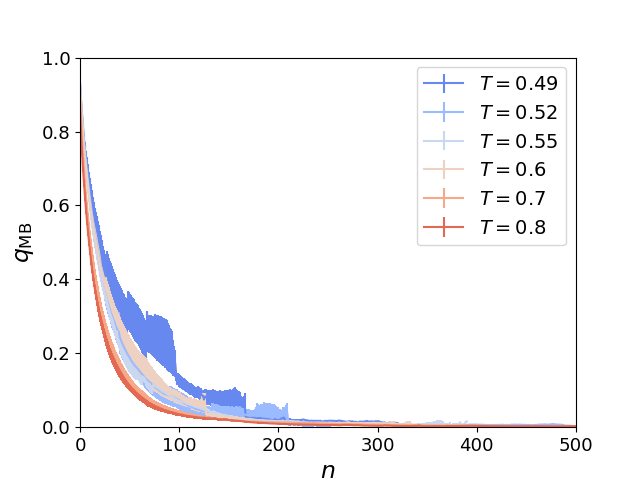}
 \caption{Same as Fig.~\ref{fig:qMB}, but with a different operational definition of MBs, from Ref.~\cite{doliwa:03c}. }
 \label{fig:qMB-heuer}
\end{figure}

A different, though similar, procedure for the identification of the MBs from the trajectory is provided in Ref.~\cite{doliwa:03c}. This procedure consists of (a) identifying all the time intervals in which each IS appears; (b) separating or (c) merging partially overlapping time intervals according to heuristic criteria; (d) removing intervals that are completely contained in other intervals; (e) stating that each of the remaining intervals is a MB, with $\tmb$ corresponding to the length of the interval; (f) assigning to each MB the lowest energy that is visited during that time interval.

We repeated our analyses using this definition of MB, and obtained analogous results. 
The MB times are slightly higher, as well as the barriers, and $\qmb$ decays faster (Fig.~\ref{fig:qMB-heuer}),
but the overall qualitative picture does not change.

\bibliographystyle{apsrev4-2}
\bibliography{marco}

\begin{thebibliography}{64}%
\makeatletter
\providecommand \@ifxundefined [1]{%
 \@ifx{#1\undefined}
}%
\providecommand \@ifnum [1]{%
 \ifnum #1\expandafter \@firstoftwo
 \else \expandafter \@secondoftwo
 \fi
}%
\providecommand \@ifx [1]{%
 \ifx #1\expandafter \@firstoftwo
 \else \expandafter \@secondoftwo
 \fi
}%
\providecommand \natexlab [1]{#1}%
\providecommand \enquote  [1]{``#1''}%
\providecommand \bibnamefont  [1]{#1}%
\providecommand \bibfnamefont [1]{#1}%
\providecommand \citenamefont [1]{#1}%
\providecommand \href@noop [0]{\@secondoftwo}%
\providecommand \href [0]{\begingroup \@sanitize@url \@href}%
\providecommand \@href[1]{\@@startlink{#1}\@@href}%
\providecommand \@@href[1]{\endgroup#1\@@endlink}%
\providecommand \@sanitize@url [0]{\catcode `\\12\catcode `\$12\catcode
  `\&12\catcode `\#12\catcode `\^12\catcode `\_12\catcode `\%12\relax}%
\providecommand \@@startlink[1]{}%
\providecommand \@@endlink[0]{}%
\providecommand \url  [0]{\begingroup\@sanitize@url \@url }%
\providecommand \@url [1]{\endgroup\@href {#1}{\urlprefix }}%
\providecommand \urlprefix  [0]{URL }%
\providecommand \Eprint [0]{\href }%
\providecommand \doibase [0]{https://doi.org/}%
\providecommand \selectlanguage [0]{\@gobble}%
\providecommand \bibinfo  [0]{\@secondoftwo}%
\providecommand \bibfield  [0]{\@secondoftwo}%
\providecommand \translation [1]{[#1]}%
\providecommand \BibitemOpen [0]{}%
\providecommand \bibitemStop [0]{}%
\providecommand \bibitemNoStop [0]{.\EOS\space}%
\providecommand \EOS [0]{\spacefactor3000\relax}%
\providecommand \BibitemShut  [1]{\csname bibitem#1\endcsname}%
\let\auto@bib@innerbib\@empty
\bibitem [{\citenamefont {Cavagna}(2009)}]{cavagna:09}%
  \BibitemOpen
  \bibfield  {author} {\bibinfo {author} {\bibfnamefont {A.}~\bibnamefont
  {Cavagna}},\ }\href
  {https://doi.org/https://doi.org/10.1016/j.physrep.2009.03.003} {\bibfield
  {journal} {\bibinfo  {journal} {Phys. Rep.}\ }\textbf {\bibinfo {volume}
  {476}},\ \bibinfo {pages} {51 } (\bibinfo {year} {2009})},\ \Eprint
  {https://arxiv.org/abs/arXiv:0903.4264} {arXiv:0903.4264} \BibitemShut
  {NoStop}%
\bibitem [{\citenamefont {Altieri}\ \emph {et~al.}(2020)\citenamefont
  {Altieri}, \citenamefont {Biroli},\ and\ \citenamefont
  {Cammarota}}]{altieri:20}%
  \BibitemOpen
  \bibfield  {author} {\bibinfo {author} {\bibfnamefont {A.}~\bibnamefont
  {Altieri}}, \bibinfo {author} {\bibfnamefont {G.}~\bibnamefont {Biroli}},\
  and\ \bibinfo {author} {\bibfnamefont {C.}~\bibnamefont {Cammarota}},\
  }\href@noop {} {\bibfield  {journal} {\bibinfo  {journal} {Journal of Physics
  A: Mathematical and Theoretical}\ }\textbf {\bibinfo {volume} {53}},\
  \bibinfo {pages} {375006} (\bibinfo {year} {2020})}\BibitemShut {NoStop}%
\bibitem [{\citenamefont {Arceri}\ \emph {et~al.}(2020)\citenamefont {Arceri},
  \citenamefont {Landes}, \citenamefont {Berthier},\ and\ \citenamefont
  {Biroli}}]{arceri:20}%
  \BibitemOpen
  \bibfield  {author} {\bibinfo {author} {\bibfnamefont {F.}~\bibnamefont
  {Arceri}}, \bibinfo {author} {\bibfnamefont {F.~P.}\ \bibnamefont {Landes}},
  \bibinfo {author} {\bibfnamefont {L.}~\bibnamefont {Berthier}},\ and\
  \bibinfo {author} {\bibfnamefont {G.}~\bibnamefont {Biroli}},\ }\Eprint
  {https://arxiv.org/abs/arXiv:2006.09725} {arXiv:2006.09725}  (\bibinfo {year}
  {2020})\BibitemShut {NoStop}%
\bibitem [{\citenamefont {Goldstein}(1969)}]{goldstein:69}%
  \BibitemOpen
  \bibfield  {author} {\bibinfo {author} {\bibfnamefont {M.}~\bibnamefont
  {Goldstein}},\ }\href@noop {} {\bibfield  {journal} {\bibinfo  {journal} {The
  Journal of Chemical Physics}\ }\textbf {\bibinfo {volume} {51}},\ \bibinfo
  {pages} {3728} (\bibinfo {year} {1969})}\BibitemShut {NoStop}%
\bibitem [{\citenamefont {Sastry}\ \emph {et~al.}(1998)\citenamefont {Sastry},
  \citenamefont {Debenedetti},\ and\ \citenamefont {Stillinger}}]{sastry:98}%
  \BibitemOpen
  \bibfield  {author} {\bibinfo {author} {\bibfnamefont {S.}~\bibnamefont
  {Sastry}}, \bibinfo {author} {\bibfnamefont {P.~G.}\ \bibnamefont
  {Debenedetti}},\ and\ \bibinfo {author} {\bibfnamefont {F.~H.}\ \bibnamefont
  {Stillinger}},\ }\href@noop {} {\bibfield  {journal} {\bibinfo  {journal}
  {Nature}\ }\textbf {\bibinfo {volume} {393}},\ \bibinfo {pages} {554}
  (\bibinfo {year} {1998})}\BibitemShut {NoStop}%
\bibitem [{\citenamefont {Schr\o{}der}\ \emph {et~al.}(2000)\citenamefont
  {Schr\o{}der}, \citenamefont {Sastry}, \citenamefont {Dyre},\ and\
  \citenamefont {Glotzer}}]{schroder:00}%
  \BibitemOpen
  \bibfield  {author} {\bibinfo {author} {\bibfnamefont {T.~B.}\ \bibnamefont
  {Schr\o{}der}}, \bibinfo {author} {\bibfnamefont {S.}~\bibnamefont {Sastry}},
  \bibinfo {author} {\bibfnamefont {J.~C.}\ \bibnamefont {Dyre}},\ and\
  \bibinfo {author} {\bibfnamefont {S.~C.}\ \bibnamefont {Glotzer}},\ }\href
  {https://doi.org/10.1063/1.481621} {\bibfield  {journal} {\bibinfo  {journal}
  {The Journal of Chemical Physics}\ }\textbf {\bibinfo {volume} {112}},\
  \bibinfo {pages} {9834} (\bibinfo {year} {2000})},\ \Eprint
  {https://arxiv.org/abs/https://doi.org/10.1063/1.481621}
  {https://doi.org/10.1063/1.481621} \BibitemShut {NoStop}%
\bibitem [{\citenamefont {Dzero}\ \emph {et~al.}(2005)\citenamefont {Dzero},
  \citenamefont {Schmalian},\ and\ \citenamefont {Wolynes}}]{dzero:05}%
  \BibitemOpen
  \bibfield  {author} {\bibinfo {author} {\bibfnamefont {M.}~\bibnamefont
  {Dzero}}, \bibinfo {author} {\bibfnamefont {J.}~\bibnamefont {Schmalian}},\
  and\ \bibinfo {author} {\bibfnamefont {P.~G.}\ \bibnamefont {Wolynes}},\
  }\href {https://doi.org/10.1103/PhysRevB.72.100201} {\bibfield  {journal}
  {\bibinfo  {journal} {Phys. Rev. B}\ }\textbf {\bibinfo {volume} {72}},\
  \bibinfo {pages} {100201} (\bibinfo {year} {2005})}\BibitemShut {NoStop}%
\bibitem [{\citenamefont {Arrhenius}(1889)}]{arrhenius:1889}%
  \BibitemOpen
  \bibfield  {author} {\bibinfo {author} {\bibfnamefont {S.}~\bibnamefont
  {Arrhenius}},\ }\href {https://doi.org/doi:10.1515/zpch-1889-0108} {\bibfield
   {journal} {\bibinfo  {journal} {Z. Phys. Chem.}\ }\textbf {\bibinfo {volume}
  {4}},\ \bibinfo {pages} {96} (\bibinfo {year} {1889})}\BibitemShut {NoStop}%
\bibitem [{\citenamefont {Dyre}(1987)}]{dyre:87}%
  \BibitemOpen
  \bibfield  {author} {\bibinfo {author} {\bibfnamefont {J.~C.}\ \bibnamefont
  {Dyre}},\ }\href {https://doi.org/10.1103/PhysRevLett.58.792} {\bibfield
  {journal} {\bibinfo  {journal} {Phys. Rev. Lett.}\ }\textbf {\bibinfo
  {volume} {58}},\ \bibinfo {pages} {792} (\bibinfo {year} {1987})}\BibitemShut
  {NoStop}%
\bibitem [{\citenamefont {Bouchaud}(1992)}]{bouchaud:92}%
  \BibitemOpen
  \bibfield  {author} {\bibinfo {author} {\bibfnamefont {J.}~\bibnamefont
  {Bouchaud}},\ }\href {https://doi.org/10.1051/jp1:1992238} {\bibfield
  {journal} {\bibinfo  {journal} {J. Phys. I France}\ }\textbf {\bibinfo
  {volume} {2}},\ \bibinfo {pages} {1705} (\bibinfo {year} {1992})}\BibitemShut
  {NoStop}%
\bibitem [{\citenamefont {Monthus}\ and\ \citenamefont
  {Bouchaud}(1996)}]{monthus:96}%
  \BibitemOpen
  \bibfield  {author} {\bibinfo {author} {\bibfnamefont {C.}~\bibnamefont
  {Monthus}}\ and\ \bibinfo {author} {\bibfnamefont {J.}~\bibnamefont
  {Bouchaud}},\ }\href
  {https://doi.org/https://doi.org/10.1088/0305-4470/29/14/012} {\bibfield
  {journal} {\bibinfo  {journal} {J. Phys. A}\ }\textbf {\bibinfo {volume}
  {29}},\ \bibinfo {pages} {3847} (\bibinfo {year} {1996})}\BibitemShut
  {NoStop}%
\bibitem [{\citenamefont {Denny}\ \emph {et~al.}(2003)\citenamefont {Denny},
  \citenamefont {Reichman},\ and\ \citenamefont {Bouchaud}}]{denny:03}%
  \BibitemOpen
  \bibfield  {author} {\bibinfo {author} {\bibfnamefont {R.~A.}\ \bibnamefont
  {Denny}}, \bibinfo {author} {\bibfnamefont {D.~R.}\ \bibnamefont
  {Reichman}},\ and\ \bibinfo {author} {\bibfnamefont {J.-P.}\ \bibnamefont
  {Bouchaud}},\ }\href {https://doi.org/10.1103/PhysRevLett.90.025503}
  {\bibfield  {journal} {\bibinfo  {journal} {Phys. Rev. Lett.}\ }\textbf
  {\bibinfo {volume} {90}},\ \bibinfo {pages} {025503} (\bibinfo {year}
  {2003})}\BibitemShut {NoStop}%
\bibitem [{\citenamefont {Gayrard}(2018)}]{gayrard:18}%
  \BibitemOpen
  \bibfield  {author} {\bibinfo {author} {\bibfnamefont {V.}~\bibnamefont
  {Gayrard}},\ }\bibfield  {journal} {\bibinfo  {journal} {Probab. Theory
  Relat. Fields}\ }\href {https://doi.org/10.1007/s00440-018-0873-6}
  {10.1007/s00440-018-0873-6} (\bibinfo {year} {2018}),\ \Eprint
  {https://arxiv.org/abs/arXiv:1602.06081} {arXiv:1602.06081} \BibitemShut
  {NoStop}%
\bibitem [{\citenamefont {Baity-Jesi}\ \emph
  {et~al.}(2018{\natexlab{a}})\citenamefont {Baity-Jesi}, \citenamefont
  {Biroli},\ and\ \citenamefont {Cammarota}}]{baityjesi:18}%
  \BibitemOpen
  \bibfield  {author} {\bibinfo {author} {\bibfnamefont {M.}~\bibnamefont
  {Baity-Jesi}}, \bibinfo {author} {\bibfnamefont {G.}~\bibnamefont {Biroli}},\
  and\ \bibinfo {author} {\bibfnamefont {C.}~\bibnamefont {Cammarota}},\ }\href
  {http://stacks.iop.org/1742-5468/2018/i=1/a=013301} {\bibfield  {journal}
  {\bibinfo  {journal} {J. Stat. Mech.: Theory Exp}\ ,\ \bibinfo {pages}
  {013301}} (\bibinfo {year} {2018}{\natexlab{a}})},\ \Eprint
  {https://arxiv.org/abs/arXiv:1708.03268} {arXiv:1708.03268} \BibitemShut
  {NoStop}%
\bibitem [{\citenamefont {Ben-Arous}\ \emph {et~al.}(2008)\citenamefont
  {Ben-Arous}, \citenamefont {Bovier},\ and\ \citenamefont
  {{\v{C}}ern{\`y}}}]{benarous:08}%
  \BibitemOpen
  \bibfield  {author} {\bibinfo {author} {\bibfnamefont {G.}~\bibnamefont
  {Ben-Arous}}, \bibinfo {author} {\bibfnamefont {A.}~\bibnamefont {Bovier}},\
  and\ \bibinfo {author} {\bibfnamefont {J.}~\bibnamefont {{\v{C}}ern{\`y}}},\
  }\href {https://doi.org/10.1007/s00220-008-0565-7} {\bibfield  {journal}
  {\bibinfo  {journal} {Communications in Mathematical Physics}\ }\textbf
  {\bibinfo {volume} {282}},\ \bibinfo {pages} {663} (\bibinfo {year}
  {2008})}\BibitemShut {NoStop}%
\bibitem [{\citenamefont {Gayrard}(2016)}]{gayrard:16}%
  \BibitemOpen
  \bibfield  {author} {\bibinfo {author} {\bibfnamefont {V.}~\bibnamefont
  {Gayrard}},\ }\href {https://doi.org/10.1007/s00023-015-0442-9} {\bibfield
  {journal} {\bibinfo  {journal} {Annales Henri Poincar{\'e}}\ }\textbf
  {\bibinfo {volume} {17}},\ \bibinfo {pages} {537} (\bibinfo {year}
  {2016})}\BibitemShut {NoStop}%
\bibitem [{\citenamefont {Baity-Jesi}\ \emph
  {et~al.}(2018{\natexlab{b}})\citenamefont {Baity-Jesi}, \citenamefont
  {Achard-de Lustrac},\ and\ \citenamefont {Biroli}}]{baityjesi:18c}%
  \BibitemOpen
  \bibfield  {author} {\bibinfo {author} {\bibfnamefont {M.}~\bibnamefont
  {Baity-Jesi}}, \bibinfo {author} {\bibfnamefont {A.}~\bibnamefont {Achard-de
  Lustrac}},\ and\ \bibinfo {author} {\bibfnamefont {G.}~\bibnamefont
  {Biroli}},\ }\href@noop {} {\bibfield  {journal} {\bibinfo  {journal} {Phys.
  Rev. E}\ }\textbf {\bibinfo {volume} {98}},\ \bibinfo {pages} {012133}
  (\bibinfo {year} {2018}{\natexlab{b}})},\ \Eprint
  {https://arxiv.org/abs/arXiv:1805.04581} {arXiv:1805.04581} \BibitemShut
  {NoStop}%
\bibitem [{\citenamefont {Stariolo}\ and\ \citenamefont
  {Cugliandolo}(2019)}]{stariolo:19}%
  \BibitemOpen
  \bibfield  {author} {\bibinfo {author} {\bibfnamefont {D.~A.}\ \bibnamefont
  {Stariolo}}\ and\ \bibinfo {author} {\bibfnamefont {L.~F.}\ \bibnamefont
  {Cugliandolo}},\ }\href {https://doi.org/10.1209/0295-5075/127/16002}
  {\bibfield  {journal} {\bibinfo  {journal} {{EPL} (Europhysics Letters)}\
  }\textbf {\bibinfo {volume} {127}},\ \bibinfo {pages} {16002} (\bibinfo
  {year} {2019})},\ \Eprint {https://arxiv.org/abs/arXiv:1904.10731}
  {arXiv:1904.10731} \BibitemShut {NoStop}%
\bibitem [{\citenamefont {Stariolo}\ and\ \citenamefont
  {Cugliandolo}(2020)}]{stariolo:20}%
  \BibitemOpen
  \bibfield  {author} {\bibinfo {author} {\bibfnamefont {D.~A.}\ \bibnamefont
  {Stariolo}}\ and\ \bibinfo {author} {\bibfnamefont {L.~F.}\ \bibnamefont
  {Cugliandolo}},\ }\href {https://doi.org/10.1103/PhysRevE.102.022126}
  {\bibfield  {journal} {\bibinfo  {journal} {Phys. Rev. E}\ }\textbf {\bibinfo
  {volume} {102}},\ \bibinfo {pages} {022126} (\bibinfo {year} {2020})},\
  \Eprint {https://arxiv.org/abs/arXiv:2004.09410} {arXiv:2004.09410}
  \BibitemShut {NoStop}%
\bibitem [{\citenamefont {Angelani}\ \emph {et~al.}(2000)\citenamefont
  {Angelani}, \citenamefont {Di~Leonardo}, \citenamefont {Ruocco},
  \citenamefont {Scala},\ and\ \citenamefont {Sciortino}}]{angelani:00}%
  \BibitemOpen
  \bibfield  {author} {\bibinfo {author} {\bibfnamefont {L.}~\bibnamefont
  {Angelani}}, \bibinfo {author} {\bibfnamefont {R.}~\bibnamefont
  {Di~Leonardo}}, \bibinfo {author} {\bibfnamefont {G.}~\bibnamefont {Ruocco}},
  \bibinfo {author} {\bibfnamefont {A.}~\bibnamefont {Scala}},\ and\ \bibinfo
  {author} {\bibfnamefont {F.}~\bibnamefont {Sciortino}},\ }\href
  {https://doi.org/10.1103/PhysRevLett.85.5356} {\bibfield  {journal} {\bibinfo
   {journal} {Phys. Rev. Lett.}\ }\textbf {\bibinfo {volume} {85}},\ \bibinfo
  {pages} {5356} (\bibinfo {year} {2000})}\BibitemShut {NoStop}%
\bibitem [{\citenamefont {Broderix}\ \emph {et~al.}(2000)\citenamefont
  {Broderix}, \citenamefont {Bhattacharya}, \citenamefont {Cavagna},
  \citenamefont {Zippelius},\ and\ \citenamefont {Giardina}}]{broderix:00}%
  \BibitemOpen
  \bibfield  {author} {\bibinfo {author} {\bibfnamefont {K.}~\bibnamefont
  {Broderix}}, \bibinfo {author} {\bibfnamefont {K.~K.}\ \bibnamefont
  {Bhattacharya}}, \bibinfo {author} {\bibfnamefont {A.}~\bibnamefont
  {Cavagna}}, \bibinfo {author} {\bibfnamefont {A.}~\bibnamefont {Zippelius}},\
  and\ \bibinfo {author} {\bibfnamefont {I.}~\bibnamefont {Giardina}},\ }\href
  {https://doi.org/10.1103/PhysRevLett.85.5360} {\bibfield  {journal} {\bibinfo
   {journal} {Phys. Rev. Lett.}\ }\textbf {\bibinfo {volume} {85}},\ \bibinfo
  {pages} {5360} (\bibinfo {year} {2000})}\BibitemShut {NoStop}%
\bibitem [{\citenamefont {Heuer}(2008)}]{heuer:08}%
  \BibitemOpen
  \bibfield  {author} {\bibinfo {author} {\bibfnamefont {A.}~\bibnamefont
  {Heuer}},\ }\href@noop {} {\bibfield  {journal} {\bibinfo  {journal} {J.
  Phys. Condens. Matter}\ }\textbf {\bibinfo {volume} {20}},\ \bibinfo {pages}
  {373101} (\bibinfo {year} {2008})}\BibitemShut {NoStop}%
\bibitem [{\citenamefont {Doliwa}\ and\ \citenamefont
  {Heuer}(2003{\natexlab{a}})}]{doliwa:03d}%
  \BibitemOpen
  \bibfield  {author} {\bibinfo {author} {\bibfnamefont {B.}~\bibnamefont
  {Doliwa}}\ and\ \bibinfo {author} {\bibfnamefont {A.}~\bibnamefont {Heuer}},\
  }\href@noop {} {\bibfield  {journal} {\bibinfo  {journal} {Physical Review
  E}\ }\textbf {\bibinfo {volume} {67}},\ \bibinfo {pages} {030501} (\bibinfo
  {year} {2003}{\natexlab{a}})}\BibitemShut {NoStop}%
\bibitem [{\citenamefont {Doliwa}\ and\ \citenamefont
  {Heuer}(2003{\natexlab{b}})}]{doliwa:03}%
  \BibitemOpen
  \bibfield  {author} {\bibinfo {author} {\bibfnamefont {B.}~\bibnamefont
  {Doliwa}}\ and\ \bibinfo {author} {\bibfnamefont {A.}~\bibnamefont {Heuer}},\
  }\href {https://doi.org/10.1103/PhysRevLett.91.235501} {\bibfield  {journal}
  {\bibinfo  {journal} {Phys. Rev. Lett.}\ }\textbf {\bibinfo {volume} {91}},\
  \bibinfo {pages} {235501} (\bibinfo {year} {2003}{\natexlab{b}})}\BibitemShut
  {NoStop}%
\bibitem [{\citenamefont {Doliwa}\ and\ \citenamefont
  {Heuer}(2003{\natexlab{c}})}]{doliwa:03c}%
  \BibitemOpen
  \bibfield  {author} {\bibinfo {author} {\bibfnamefont {B.}~\bibnamefont
  {Doliwa}}\ and\ \bibinfo {author} {\bibfnamefont {A.}~\bibnamefont {Heuer}},\
  }\href {https://arxiv.org/pdf/cond-mat/0209139.pdf} {\bibfield  {journal}
  {\bibinfo  {journal} {Physical Review E}\ }\textbf {\bibinfo {volume} {67}},\
  \bibinfo {pages} {031506} (\bibinfo {year} {2003}{\natexlab{c}})},\ \Eprint
  {https://arxiv.org/abs/arxiv:cond-mat/0209139} {arxiv:cond-mat/0209139}
  \BibitemShut {NoStop}%
\bibitem [{\citenamefont {Fris}\ \emph {et~al.}(2011)\citenamefont {Fris},
  \citenamefont {Appignanesi},\ and\ \citenamefont {Weeks}}]{fris:11}%
  \BibitemOpen
  \bibfield  {author} {\bibinfo {author} {\bibfnamefont {J.~A.~R.}\
  \bibnamefont {Fris}}, \bibinfo {author} {\bibfnamefont {G.~A.}\ \bibnamefont
  {Appignanesi}},\ and\ \bibinfo {author} {\bibfnamefont {E.~R.}\ \bibnamefont
  {Weeks}},\ }\href@noop {} {\bibfield  {journal} {\bibinfo  {journal} {Phys.
  Rev. Lett}\ }\textbf {\bibinfo {volume} {107}},\ \bibinfo {pages} {065704}
  (\bibinfo {year} {2011})}\BibitemShut {NoStop}%
\bibitem [{\citenamefont {Rizzo}(2020)}]{rizzo:20}%
  \BibitemOpen
  \bibfield  {author} {\bibinfo {author} {\bibfnamefont {T.}~\bibnamefont
  {Rizzo}},\ }\Eprint {https://arxiv.org/abs/arXiv:2012.09556}
  {arXiv:2012.09556}  (\bibinfo {year} {2020})\BibitemShut {NoStop}%
\bibitem [{\citenamefont {Heuer}\ \emph {et~al.}(2005)\citenamefont {Heuer},
  \citenamefont {Doliwa},\ and\ \citenamefont {Saksaengwijit}}]{heuer:05}%
  \BibitemOpen
  \bibfield  {author} {\bibinfo {author} {\bibfnamefont {A.}~\bibnamefont
  {Heuer}}, \bibinfo {author} {\bibfnamefont {B.}~\bibnamefont {Doliwa}},\ and\
  \bibinfo {author} {\bibfnamefont {A.}~\bibnamefont {Saksaengwijit}},\ }\href
  {https://doi.org/https://doi.org/10.1103/PhysRevE.72.021503} {\bibfield
  {journal} {\bibinfo  {journal} {Physical Review E}\ }\textbf {\bibinfo
  {volume} {72}},\ \bibinfo {pages} {021503} (\bibinfo {year}
  {2005})}\BibitemShut {NoStop}%
\bibitem [{\citenamefont {Berthier}\ and\ \citenamefont
  {Garrahan}(2003)}]{berthier:03b}%
  \BibitemOpen
  \bibfield  {author} {\bibinfo {author} {\bibfnamefont {L.}~\bibnamefont
  {Berthier}}\ and\ \bibinfo {author} {\bibfnamefont {J.~P.}\ \bibnamefont
  {Garrahan}},\ }\href {https://doi.org/10.1063/1.1593020} {\bibfield
  {journal} {\bibinfo  {journal} {The Journal of Chemical Physics}\ }\textbf
  {\bibinfo {volume} {119}},\ \bibinfo {pages} {4367} (\bibinfo {year}
  {2003})}\BibitemShut {NoStop}%
\bibitem [{\citenamefont {Berthier}\ and\ \citenamefont
  {Garrahan}(2005)}]{berthier:05c}%
  \BibitemOpen
  \bibfield  {author} {\bibinfo {author} {\bibfnamefont {L.}~\bibnamefont
  {Berthier}}\ and\ \bibinfo {author} {\bibfnamefont {J.~P.}\ \bibnamefont
  {Garrahan}},\ }\href@noop {} {\bibfield  {journal} {\bibinfo  {journal} {The
  journal of physical chemistry B}\ }\textbf {\bibinfo {volume} {109}},\
  \bibinfo {pages} {3578} (\bibinfo {year} {2005})}\BibitemShut {NoStop}%
\bibitem [{\citenamefont {Kob}\ and\ \citenamefont {Andersen}(1994)}]{kob:94}%
  \BibitemOpen
  \bibfield  {author} {\bibinfo {author} {\bibfnamefont {W.}~\bibnamefont
  {Kob}}\ and\ \bibinfo {author} {\bibfnamefont {H.~C.}\ \bibnamefont
  {Andersen}},\ }\href@noop {} {\bibfield  {journal} {\bibinfo  {journal}
  {Phys. Rev. Lett.}\ }\textbf {\bibinfo {volume} {73}},\ \bibinfo {pages}
  {1376} (\bibinfo {year} {1994})}\BibitemShut {NoStop}%
\bibitem [{\citenamefont {Doliwa}\ and\ \citenamefont
  {Heuer}(2003{\natexlab{d}})}]{doliwa:03b}%
  \BibitemOpen
  \bibfield  {author} {\bibinfo {author} {\bibfnamefont {B.}~\bibnamefont
  {Doliwa}}\ and\ \bibinfo {author} {\bibfnamefont {A.}~\bibnamefont {Heuer}},\
  }\href {http://stacks.iop.org/0953-8984/15/i=11/a=309} {\bibfield  {journal}
  {\bibinfo  {journal} {J. Phys. Condens. Matter}\ }\textbf {\bibinfo {volume}
  {15}},\ \bibinfo {pages} {S849} (\bibinfo {year} {2003}{\natexlab{d}})},\
  \Eprint {https://arxiv.org/abs/arXiv:cond-mat/0210121}
  {arXiv:cond-mat/0210121} \BibitemShut {NoStop}%
\bibitem [{\citenamefont {Anderson}\ \emph {et~al.}(2008)\citenamefont
  {Anderson}, \citenamefont {Lorenz},\ and\ \citenamefont
  {Travesset}}]{anderson:08}%
  \BibitemOpen
  \bibfield  {author} {\bibinfo {author} {\bibfnamefont {J.~A.}\ \bibnamefont
  {Anderson}}, \bibinfo {author} {\bibfnamefont {C.~D.}\ \bibnamefont
  {Lorenz}},\ and\ \bibinfo {author} {\bibfnamefont {A.}~\bibnamefont
  {Travesset}},\ }\href {https://doi.org/10.1016/j.jcp.2008.01.047} {\bibfield
  {journal} {\bibinfo  {journal} {Journal of Computational Physics}\ }\textbf
  {\bibinfo {volume} {227}},\ \bibinfo {pages} {5342} (\bibinfo {year}
  {2008})},\ \bibinfo {note} {{HOOMD}-blue feature: {HOOMD}-blue}\BibitemShut
  {NoStop}%
\bibitem [{\citenamefont {Glaser}\ \emph {et~al.}(2015)\citenamefont {Glaser},
  \citenamefont {Nguyen}, \citenamefont {Anderson}, \citenamefont {Liu},
  \citenamefont {Spiga}, \citenamefont {Millan}, \citenamefont {Morse},\ and\
  \citenamefont {Glotzer}}]{glaser:15}%
  \BibitemOpen
  \bibfield  {author} {\bibinfo {author} {\bibfnamefont {J.}~\bibnamefont
  {Glaser}}, \bibinfo {author} {\bibfnamefont {T.~D.}\ \bibnamefont {Nguyen}},
  \bibinfo {author} {\bibfnamefont {J.~A.}\ \bibnamefont {Anderson}}, \bibinfo
  {author} {\bibfnamefont {P.}~\bibnamefont {Liu}}, \bibinfo {author}
  {\bibfnamefont {F.}~\bibnamefont {Spiga}}, \bibinfo {author} {\bibfnamefont
  {J.~A.}\ \bibnamefont {Millan}}, \bibinfo {author} {\bibfnamefont {D.~C.}\
  \bibnamefont {Morse}},\ and\ \bibinfo {author} {\bibfnamefont {S.~C.}\
  \bibnamefont {Glotzer}},\ }\href {https://doi.org/10.1016/j.cpc.2015.02.028}
  {\bibfield  {journal} {\bibinfo  {journal} {Computer Physics Communications}\
  }\textbf {\bibinfo {volume} {192}},\ \bibinfo {pages} {97} (\bibinfo {year}
  {2015})},\ \bibinfo {note} {{HOOMD}-blue feature: {HOOMD}-blue}\BibitemShut
  {NoStop}%
\bibitem [{cod()}]{code:structuralglass}%
  \BibitemOpen
  \href@noop {} {}\bibinfo {note} {The code used for all our simulations and
  analyses is available at
  \texttt{github.com/mbaityje/STRUCTURAL-GLASS}}\BibitemShut {NoStop}%
\bibitem [{\citenamefont {Karmakar}\ \emph {et~al.}(2009)\citenamefont
  {Karmakar}, \citenamefont {Dasgupta},\ and\ \citenamefont
  {Sastry}}]{karmakar:09}%
  \BibitemOpen
  \bibfield  {author} {\bibinfo {author} {\bibfnamefont {S.}~\bibnamefont
  {Karmakar}}, \bibinfo {author} {\bibfnamefont {C.}~\bibnamefont {Dasgupta}},\
  and\ \bibinfo {author} {\bibfnamefont {S.}~\bibnamefont {Sastry}},\
  }\href@noop {} {\bibfield  {journal} {\bibinfo  {journal} {Proceedings of the
  National Academy of Sciences}\ }\textbf {\bibinfo {volume} {106}},\ \bibinfo
  {pages} {3675} (\bibinfo {year} {2009})}\BibitemShut {NoStop}%
\bibitem [{\citenamefont {Folena}\ \emph {et~al.}(2020)\citenamefont {Folena},
  \citenamefont {Franz},\ and\ \citenamefont {Ricci-Tersenghi}}]{folena:20}%
  \BibitemOpen
  \bibfield  {author} {\bibinfo {author} {\bibfnamefont {G.}~\bibnamefont
  {Folena}}, \bibinfo {author} {\bibfnamefont {S.}~\bibnamefont {Franz}},\ and\
  \bibinfo {author} {\bibfnamefont {F.}~\bibnamefont {Ricci-Tersenghi}},\
  }\href {https://doi.org/10.1103/PhysRevX.10.031045} {\bibfield  {journal}
  {\bibinfo  {journal} {Phys. Rev. X}\ }\textbf {\bibinfo {volume} {10}},\
  \bibinfo {pages} {031045} (\bibinfo {year} {2020})},\ \Eprint
  {https://arxiv.org/abs/arXiv:1903.01421} {arXiv:1903.01421} \BibitemShut
  {NoStop}%
\bibitem [{\citenamefont {Baity-Jesi}\ and\ \citenamefont
  {Mart\'in-Mayor}(2019)}]{baityjesi:19b}%
  \BibitemOpen
  \bibfield  {author} {\bibinfo {author} {\bibfnamefont {M.}~\bibnamefont
  {Baity-Jesi}}\ and\ \bibinfo {author} {\bibfnamefont {V.}~\bibnamefont
  {Mart\'in-Mayor}},\ }\href@noop {} {\bibfield  {journal} {\bibinfo  {journal}
  {Journal of Statistical Mechanics: Theory and Experiment}\ }\textbf {\bibinfo
  {volume} {2019}},\ \bibinfo {pages} {084016} (\bibinfo {year} {2019})},\
  \Eprint {https://arxiv.org/abs/arXiv:1901.05581} {arXiv:1901.05581}
  \BibitemShut {NoStop}%
\bibitem [{\citenamefont {Berthier}(2020)}]{berthier:20}%
  \BibitemOpen
  \bibfield  {author} {\bibinfo {author} {\bibfnamefont {L.}~\bibnamefont
  {Berthier}},\ }\Eprint {https://arxiv.org/abs/arXiv:2010.12244}
  {arXiv:2010.12244}  (\bibinfo {year} {2020})\BibitemShut {NoStop}%
\bibitem [{\citenamefont {Rizzo}(2013)}]{rizzo:13}%
  \BibitemOpen
  \bibfield  {author} {\bibinfo {author} {\bibfnamefont {T.}~\bibnamefont
  {Rizzo}},\ }\href {https://doi.org/10.1103/PhysRevE.88.032135} {\bibfield
  {journal} {\bibinfo  {journal} {Phys. Rev. E}\ }\textbf {\bibinfo {volume}
  {88}},\ \bibinfo {pages} {032135} (\bibinfo {year} {2013})}\BibitemShut
  {NoStop}%
\bibitem [{\citenamefont {Barrat}\ and\ \citenamefont
  {M{\'e}zard}(1995)}]{barrat:95}%
  \BibitemOpen
  \bibfield  {author} {\bibinfo {author} {\bibfnamefont {A.}~\bibnamefont
  {Barrat}}\ and\ \bibinfo {author} {\bibfnamefont {M.}~\bibnamefont
  {M{\'e}zard}},\ }\href@noop {} {\bibfield  {journal} {\bibinfo  {journal}
  {Journal de Physique I}\ }\textbf {\bibinfo {volume} {5}},\ \bibinfo {pages}
  {941} (\bibinfo {year} {1995})}\BibitemShut {NoStop}%
\bibitem [{\citenamefont {Cammarota}\ and\ \citenamefont
  {Marinari}(2015)}]{cammarota:15}%
  \BibitemOpen
  \bibfield  {author} {\bibinfo {author} {\bibfnamefont {C.}~\bibnamefont
  {Cammarota}}\ and\ \bibinfo {author} {\bibfnamefont {E.}~\bibnamefont
  {Marinari}},\ }\href
  {https://doi.org/https://doi.org/10.1103/PhysRevE.92.010301} {\bibfield
  {journal} {\bibinfo  {journal} {Phys. Rev. E}\ }\textbf {\bibinfo {volume}
  {92}},\ \bibinfo {pages} {010301(R)} (\bibinfo {year} {2015})},\ \Eprint
  {https://arxiv.org/abs/arXiv:1410.2116} {arXiv:1410.2116} \BibitemShut
  {NoStop}%
\bibitem [{\citenamefont {Yaida}\ \emph {et~al.}(2016)\citenamefont {Yaida},
  \citenamefont {Berthier}, \citenamefont {Charbonneau},\ and\ \citenamefont
  {Tarjus}}]{yaida:16}%
  \BibitemOpen
  \bibfield  {author} {\bibinfo {author} {\bibfnamefont {S.}~\bibnamefont
  {Yaida}}, \bibinfo {author} {\bibfnamefont {L.}~\bibnamefont {Berthier}},
  \bibinfo {author} {\bibfnamefont {P.}~\bibnamefont {Charbonneau}},\ and\
  \bibinfo {author} {\bibfnamefont {G.}~\bibnamefont {Tarjus}},\ }\href
  {https://doi.org/10.1103/PhysRevE.94.032605} {\bibfield  {journal} {\bibinfo
  {journal} {Phys. Rev. E}\ }\textbf {\bibinfo {volume} {94}},\ \bibinfo
  {pages} {032605} (\bibinfo {year} {2016})}\BibitemShut {NoStop}%
\bibitem [{\citenamefont {J{\'o}nsson}\ \emph {et~al.}(1998)\citenamefont
  {J{\'o}nsson}, \citenamefont {Mills},\ and\ \citenamefont
  {Jacobsen}}]{jonsson:98}%
  \BibitemOpen
  \bibfield  {author} {\bibinfo {author} {\bibfnamefont {H.}~\bibnamefont
  {J{\'o}nsson}}, \bibinfo {author} {\bibfnamefont {G.}~\bibnamefont {Mills}},\
  and\ \bibinfo {author} {\bibfnamefont {K.~W.}\ \bibnamefont {Jacobsen}},\
  }in\ \href@noop {} {\emph {\bibinfo {booktitle} {Classical and quantum
  dynamics in condensed phase simulations}}}\ (\bibinfo  {publisher} {World
  Scientific},\ \bibinfo {year} {1998})\ pp.\ \bibinfo {pages}
  {385--404}\BibitemShut {NoStop}%
\bibitem [{\citenamefont {Draxler}\ \emph {et~al.}(2018)\citenamefont
  {Draxler}, \citenamefont {Veschgini}, \citenamefont {Salmhofer},\ and\
  \citenamefont {Hamprecht}}]{draxler:18}%
  \BibitemOpen
  \bibfield  {author} {\bibinfo {author} {\bibfnamefont {F.}~\bibnamefont
  {Draxler}}, \bibinfo {author} {\bibfnamefont {K.}~\bibnamefont {Veschgini}},
  \bibinfo {author} {\bibfnamefont {M.}~\bibnamefont {Salmhofer}},\ and\
  \bibinfo {author} {\bibfnamefont {F.~A.}\ \bibnamefont {Hamprecht}},\
  }\href@noop {} {\bibfield  {journal} {\bibinfo  {journal} {Proceedings of the
  35th International Conference on Machine Learning}\ }\textbf {\bibinfo
  {volume} {80}},\ \bibinfo {pages} {1308} (\bibinfo {year} {2018})},\ \Eprint
  {https://arxiv.org/abs/arXiv:1803.00885} {arXiv:1803.00885} \BibitemShut
  {NoStop}%
\bibitem [{\citenamefont {Wang}\ and\ \citenamefont
  {Stratt}(2007{\natexlab{a}})}]{wang:07}%
  \BibitemOpen
  \bibfield  {author} {\bibinfo {author} {\bibfnamefont {C.}~\bibnamefont
  {Wang}}\ and\ \bibinfo {author} {\bibfnamefont {R.~M.}\ \bibnamefont
  {Stratt}},\ }\href {https://doi.org/10.1063/1.2801994} {\bibfield  {journal}
  {\bibinfo  {journal} {The Journal of Chemical Physics}\ }\textbf {\bibinfo
  {volume} {127}},\ \bibinfo {pages} {224503} (\bibinfo {year}
  {2007}{\natexlab{a}})},\ \Eprint {https://arxiv.org/abs/arXiv:0706.4292}
  {arXiv:0706.4292} \BibitemShut {NoStop}%
\bibitem [{\citenamefont {Wang}\ and\ \citenamefont
  {Stratt}(2007{\natexlab{b}})}]{wang:07b}%
  \BibitemOpen
  \bibfield  {author} {\bibinfo {author} {\bibfnamefont {C.}~\bibnamefont
  {Wang}}\ and\ \bibinfo {author} {\bibfnamefont {R.~M.}\ \bibnamefont
  {Stratt}},\ }\href {https://doi.org/10.1063/1.2801995} {\bibfield  {journal}
  {\bibinfo  {journal} {The Journal of Chemical Physics}\ }\textbf {\bibinfo
  {volume} {127}},\ \bibinfo {pages} {224504} (\bibinfo {year}
  {2007}{\natexlab{b}})},\ \Eprint {https://arxiv.org/abs/arXiv:0706.4295}
  {arXiv:0706.4295} \BibitemShut {NoStop}%
\bibitem [{\citenamefont {Nguyen}\ \emph {et~al.}(2012)\citenamefont {Nguyen},
  \citenamefont {Isaacson}, \citenamefont {Beth~Shimmyo}, \citenamefont
  {Chen},\ and\ \citenamefont {Stratt}}]{nguyen:12}%
  \BibitemOpen
  \bibfield  {author} {\bibinfo {author} {\bibfnamefont {C.~N.}\ \bibnamefont
  {Nguyen}}, \bibinfo {author} {\bibfnamefont {J.~I.}\ \bibnamefont
  {Isaacson}}, \bibinfo {author} {\bibfnamefont {K.}~\bibnamefont
  {Beth~Shimmyo}}, \bibinfo {author} {\bibfnamefont {A.}~\bibnamefont {Chen}},\
  and\ \bibinfo {author} {\bibfnamefont {R.~M.}\ \bibnamefont {Stratt}},\
  }\href@noop {} {\bibfield  {journal} {\bibinfo  {journal} {The Journal of
  Chemical Physics}\ }\textbf {\bibinfo {volume} {136}},\ \bibinfo {pages}
  {184504} (\bibinfo {year} {2012})}\BibitemShut {NoStop}%
\bibitem [{\citenamefont {Ros}\ \emph {et~al.}(2019)\citenamefont {Ros},
  \citenamefont {Biroli},\ and\ \citenamefont {Cammarota}}]{ros2019complexity}%
  \BibitemOpen
  \bibfield  {author} {\bibinfo {author} {\bibfnamefont {V.}~\bibnamefont
  {Ros}}, \bibinfo {author} {\bibfnamefont {G.}~\bibnamefont {Biroli}},\ and\
  \bibinfo {author} {\bibfnamefont {C.}~\bibnamefont {Cammarota}},\ }\href@noop
  {} {\bibfield  {journal} {\bibinfo  {journal} {EPL (Europhysics Letters)}\
  }\textbf {\bibinfo {volume} {126}},\ \bibinfo {pages} {20003} (\bibinfo
  {year} {2019})}\BibitemShut {NoStop}%
\bibitem [{\citenamefont {Fabricius}\ and\ \citenamefont
  {Stariolo}(2004)}]{fabricius:04}%
  \BibitemOpen
  \bibfield  {author} {\bibinfo {author} {\bibfnamefont {G.}~\bibnamefont
  {Fabricius}}\ and\ \bibinfo {author} {\bibfnamefont {D.~A.}\ \bibnamefont
  {Stariolo}},\ }\href@noop {} {\bibfield  {journal} {\bibinfo  {journal}
  {Physica A: Statistical Mechanics and its Applications}\ }\textbf {\bibinfo
  {volume} {331}},\ \bibinfo {pages} {90} (\bibinfo {year} {2004})}\BibitemShut
  {NoStop}%
\bibitem [{\citenamefont {Rubner}\ and\ \citenamefont
  {Heuer}(2008)}]{rubner:08}%
  \BibitemOpen
  \bibfield  {author} {\bibinfo {author} {\bibfnamefont {O.}~\bibnamefont
  {Rubner}}\ and\ \bibinfo {author} {\bibfnamefont {A.}~\bibnamefont {Heuer}},\
  }\href {https://journals.aps.org/pre/pdf/10.1103/PhysRevE.78.011504}
  {\bibfield  {journal} {\bibinfo  {journal} {Physical Review E}\ }\textbf
  {\bibinfo {volume} {78}},\ \bibinfo {pages} {011504} (\bibinfo {year}
  {2008})}\BibitemShut {NoStop}%
\bibitem [{\citenamefont {Donati}\ \emph {et~al.}(1998)\citenamefont {Donati},
  \citenamefont {Douglas}, \citenamefont {Kob}, \citenamefont {Plimpton},
  \citenamefont {Poole},\ and\ \citenamefont {Glotzer}}]{donati:98}%
  \BibitemOpen
  \bibfield  {author} {\bibinfo {author} {\bibfnamefont {C.}~\bibnamefont
  {Donati}}, \bibinfo {author} {\bibfnamefont {J.~F.}\ \bibnamefont {Douglas}},
  \bibinfo {author} {\bibfnamefont {W.}~\bibnamefont {Kob}}, \bibinfo {author}
  {\bibfnamefont {S.~J.}\ \bibnamefont {Plimpton}}, \bibinfo {author}
  {\bibfnamefont {P.~H.}\ \bibnamefont {Poole}},\ and\ \bibinfo {author}
  {\bibfnamefont {S.~C.}\ \bibnamefont {Glotzer}},\ }\href
  {https://doi.org/10.1103/PhysRevLett.80.2338} {\bibfield  {journal} {\bibinfo
   {journal} {Phys. Rev. Lett.}\ }\textbf {\bibinfo {volume} {80}},\ \bibinfo
  {pages} {2338} (\bibinfo {year} {1998})}\BibitemShut {NoStop}%
\bibitem [{\citenamefont {Coslovich}\ \emph {et~al.}(2019)\citenamefont
  {Coslovich}, \citenamefont {Ninarello},\ and\ \citenamefont
  {Berthier}}]{coslovich:19}%
  \BibitemOpen
  \bibfield  {author} {\bibinfo {author} {\bibfnamefont {D.}~\bibnamefont
  {Coslovich}}, \bibinfo {author} {\bibfnamefont {A.}~\bibnamefont
  {Ninarello}},\ and\ \bibinfo {author} {\bibfnamefont {L.}~\bibnamefont
  {Berthier}},\ }\href@noop {} {\bibfield  {journal} {\bibinfo  {journal}
  {SciPost Physics}\ }\textbf {\bibinfo {volume} {7}},\ \bibinfo {pages} {077}
  (\bibinfo {year} {2019})}\BibitemShut {NoStop}%
\bibitem [{\citenamefont {Wyart}\ and\ \citenamefont {Cates}(2017)}]{wyart:17}%
  \BibitemOpen
  \bibfield  {author} {\bibinfo {author} {\bibfnamefont {M.}~\bibnamefont
  {Wyart}}\ and\ \bibinfo {author} {\bibfnamefont {M.~E.}\ \bibnamefont
  {Cates}},\ }\href {https://doi.org/10.1103/PhysRevLett.119.195501} {\bibfield
   {journal} {\bibinfo  {journal} {Phys. Rev. Lett.}\ }\textbf {\bibinfo
  {volume} {119}},\ \bibinfo {pages} {195501} (\bibinfo {year}
  {2017})}\BibitemShut {NoStop}%
\bibitem [{\citenamefont {Grigera}\ and\ \citenamefont
  {Parisi}(2001)}]{grigera:01}%
  \BibitemOpen
  \bibfield  {author} {\bibinfo {author} {\bibfnamefont {T.~S.}\ \bibnamefont
  {Grigera}}\ and\ \bibinfo {author} {\bibfnamefont {G.}~\bibnamefont
  {Parisi}},\ }\href@noop {} {\bibfield  {journal} {\bibinfo  {journal}
  {Physical Review E}\ }\textbf {\bibinfo {volume} {63}},\ \bibinfo {pages}
  {045102} (\bibinfo {year} {2001})}\BibitemShut {NoStop}%
\bibitem [{\citenamefont {Ninarello}\ \emph {et~al.}(2017)\citenamefont
  {Ninarello}, \citenamefont {Berthier},\ and\ \citenamefont
  {Coslovich}}]{ninarello:17}%
  \BibitemOpen
  \bibfield  {author} {\bibinfo {author} {\bibfnamefont {A.}~\bibnamefont
  {Ninarello}}, \bibinfo {author} {\bibfnamefont {L.}~\bibnamefont
  {Berthier}},\ and\ \bibinfo {author} {\bibfnamefont {D.}~\bibnamefont
  {Coslovich}},\ }\href@noop {} {\bibfield  {journal} {\bibinfo  {journal}
  {Physical Review X}\ }\textbf {\bibinfo {volume} {7}},\ \bibinfo {pages}
  {021039} (\bibinfo {year} {2017})}\BibitemShut {NoStop}%
\bibitem [{\citenamefont {Berthier}\ \emph {et~al.}(2019)\citenamefont
  {Berthier}, \citenamefont {Biroli}, \citenamefont {Bouchaud},\ and\
  \citenamefont {Tarjus}}]{berthier2019can}%
  \BibitemOpen
  \bibfield  {author} {\bibinfo {author} {\bibfnamefont {L.}~\bibnamefont
  {Berthier}}, \bibinfo {author} {\bibfnamefont {G.}~\bibnamefont {Biroli}},
  \bibinfo {author} {\bibfnamefont {J.-P.}\ \bibnamefont {Bouchaud}},\ and\
  \bibinfo {author} {\bibfnamefont {G.}~\bibnamefont {Tarjus}},\ }\href@noop {}
  {\bibfield  {journal} {\bibinfo  {journal} {The Journal of chemical physics}\
  }\textbf {\bibinfo {volume} {150}},\ \bibinfo {pages} {094501} (\bibinfo
  {year} {2019})}\BibitemShut {NoStop}%
\bibitem [{\citenamefont {Ikeda}\ \emph {et~al.}(2017)\citenamefont {Ikeda},
  \citenamefont {Zamponi},\ and\ \citenamefont {Ikeda}}]{ikeda:17}%
  \BibitemOpen
  \bibfield  {author} {\bibinfo {author} {\bibfnamefont {H.}~\bibnamefont
  {Ikeda}}, \bibinfo {author} {\bibfnamefont {F.}~\bibnamefont {Zamponi}},\
  and\ \bibinfo {author} {\bibfnamefont {A.}~\bibnamefont {Ikeda}},\
  }\href@noop {} {\bibfield  {journal} {\bibinfo  {journal} {The Journal of
  Chemical Physics}\ }\textbf {\bibinfo {volume} {147}},\ \bibinfo {pages}
  {234506} (\bibinfo {year} {2017})}\BibitemShut {NoStop}%
\bibitem [{\citenamefont {Bertin}(2003)}]{bertin:03}%
  \BibitemOpen
  \bibfield  {author} {\bibinfo {author} {\bibfnamefont {E.~M.}\ \bibnamefont
  {Bertin}},\ }\href {https://doi.org/10.1088/0305-4470/36/43/002} {\bibfield
  {journal} {\bibinfo  {journal} {J. Phys. A}\ }\textbf {\bibinfo {volume}
  {36}},\ \bibinfo {pages} {10683} (\bibinfo {year} {2003})},\ \Eprint
  {https://arxiv.org/abs/arXiv:cond-mat/0305538} {arXiv:cond-mat/0305538}
  \BibitemShut {NoStop}%
\bibitem [{\citenamefont {Cammarota}\ and\ \citenamefont
  {Marinari}(2018)}]{cammarota2018numerical}%
  \BibitemOpen
  \bibfield  {author} {\bibinfo {author} {\bibfnamefont {C.}~\bibnamefont
  {Cammarota}}\ and\ \bibinfo {author} {\bibfnamefont {E.}~\bibnamefont
  {Marinari}},\ }\href@noop {} {\bibfield  {journal} {\bibinfo  {journal}
  {Journal of Statistical Mechanics: Theory and Experiment}\ }\textbf {\bibinfo
  {volume} {2018}},\ \bibinfo {pages} {043303} (\bibinfo {year}
  {2018})}\BibitemShut {NoStop}%
\bibitem [{\citenamefont {Carbone}\ \emph {et~al.}(2020)\citenamefont
  {Carbone}, \citenamefont {Astuti},\ and\ \citenamefont
  {Baity-Jesi}}]{carbone:20}%
  \BibitemOpen
  \bibfield  {author} {\bibinfo {author} {\bibfnamefont {M.~R.}\ \bibnamefont
  {Carbone}}, \bibinfo {author} {\bibfnamefont {V.}~\bibnamefont {Astuti}},\
  and\ \bibinfo {author} {\bibfnamefont {M.}~\bibnamefont {Baity-Jesi}},\
  }\href {https://doi.org/10.1103/PhysRevE.101.052304} {\bibfield  {journal}
  {\bibinfo  {journal} {Phys. Rev. E}\ }\textbf {\bibinfo {volume} {101}},\
  \bibinfo {pages} {052304} (\bibinfo {year} {2020})},\ \Eprint
  {https://arxiv.org/abs/arXiv:2001.02567} {arXiv:2001.02567} \BibitemShut
  {NoStop}%
\bibitem [{\citenamefont {Tapias}\ \emph {et~al.}(2020)\citenamefont {Tapias},
  \citenamefont {Paprotzki},\ and\ \citenamefont {Sollich}}]{tapias:20}%
  \BibitemOpen
  \bibfield  {author} {\bibinfo {author} {\bibfnamefont {D.}~\bibnamefont
  {Tapias}}, \bibinfo {author} {\bibfnamefont {E.}~\bibnamefont {Paprotzki}},\
  and\ \bibinfo {author} {\bibfnamefont {P.}~\bibnamefont {Sollich}},\ }\href
  {https://doi.org/10.1088/1742-5468/abaecf} {\bibfield  {journal} {\bibinfo
  {journal} {Journal of Statistical Mechanics: Theory and Experiment}\ }\textbf
  {\bibinfo {volume} {2020}},\ \bibinfo {pages} {093302} (\bibinfo {year}
  {2020})},\ \Eprint {https://arxiv.org/abs/arXiv:2005.04994}
  {arXiv:2005.04994} \BibitemShut {NoStop}%
\bibitem [{\citenamefont {Fabricius}\ and\ \citenamefont
  {Stariolo}(2002)}]{fabricius:02}%
  \BibitemOpen
  \bibfield  {author} {\bibinfo {author} {\bibfnamefont {G.}~\bibnamefont
  {Fabricius}}\ and\ \bibinfo {author} {\bibfnamefont {D.~A.}\ \bibnamefont
  {Stariolo}},\ }\href {https://doi.org/10.1103/PhysRevE.66.031501} {\bibfield
  {journal} {\bibinfo  {journal} {Phys. Rev. E}\ }\textbf {\bibinfo {volume}
  {66}},\ \bibinfo {pages} {031501} (\bibinfo {year} {2002})}\BibitemShut
  {NoStop}%
\bibitem [{\citenamefont {Kirkpatrick}\ \emph {et~al.}(1989)\citenamefont
  {Kirkpatrick}, \citenamefont {Thirumalai},\ and\ \citenamefont
  {Wolynes}}]{kirkpatrick:89}%
  \BibitemOpen
  \bibfield  {author} {\bibinfo {author} {\bibfnamefont {T.~R.}\ \bibnamefont
  {Kirkpatrick}}, \bibinfo {author} {\bibfnamefont {D.}~\bibnamefont
  {Thirumalai}},\ and\ \bibinfo {author} {\bibfnamefont {P.~G.}\ \bibnamefont
  {Wolynes}},\ }\href@noop {} {\bibfield  {journal} {\bibinfo  {journal}
  {Physical Review A}\ }\textbf {\bibinfo {volume} {40}},\ \bibinfo {pages}
  {1045} (\bibinfo {year} {1989})}\BibitemShut {NoStop}%
\end{thebibliography}%

\end{document}